\documentclass[runningheads]{llncs}

\usepackage{eccv}

\usepackage{eccvabbrv}

\usepackage{graphicx}
\usepackage{booktabs}

\usepackage[accsupp]{axessibility}  

\usepackage[pagebackref,breaklinks,colorlinks,citecolor=eccvblue]{hyperref}

\usepackage{orcidlink}

\usepackage{graphicx}
\usepackage{amsmath}
\usepackage{amssymb}
\usepackage{booktabs}
\usepackage{multirow}
\usepackage{tipa}
\usepackage{algorithm}
\usepackage{algorithmic}
\usepackage{indentfirst}
\usepackage{makecell}
\usepackage{colortbl}
\usepackage{color}
\usepackage{xcolor}
\usepackage{soul}
\usepackage{url}

\usepackage{wrapfig,lipsum,booktabs}

\newcommand{\Tref}[1]{Table~\ref{#1}}

\newcommand{\Fref}[1]{Figure~\ref{#1}}
\newcommand{\Sref}[1]{Sec.~\ref{#1}}

\newcommand{\Name}{\texttt{Robust-Wide}\xspace}
\newcommand{\ModuleName}{\textit{PIDSG}\xspace}
\newcommand{\zj}[1]{\textcolor{black}{#1}}

\begin{document}

\title{\Name: Robust Watermarking against Instruction-driven Image Editing}

\titlerunning{Robust Watermarking against Instruction-driven Image Editing}

\author{Runyi Hu\inst{1,2}\orcidlink{0009-0001-6974-2542} \and
Jie Zhang\inst{1}\thanks{The corresponding author}\orcidlink{0000-0002-4230-1077} \and
Ting Xu\inst{3} \and
Jiwei Li\inst{4} \and
Tianwei Zhang\inst{1}\orcidlink{0000-0001-6595-6650}
}


\institute{Nanyang Technological University \\
\email{runyi\_hu@163.com} \\
\email{\{jie\_zhang, tianwei.zhang\}@ntu.edu.sg} \and
S-Lab, NTU \and
National University of Singapore \\
\email{xuting@nus.edu.sg} \and
Zhejiang University \\
\email{jiwei\_li@zju.edu.cn}
}

\maketitle


\begin{abstract}
Instruction-driven image editing allows users to quickly edit an image according to text instructions in a forward pass. 
Nevertheless, malicious users can easily exploit this technique to create fake images, which could cause a crisis of trust and harm the rights of the original image owners. Watermarking is a common solution to trace such malicious behavior. Unfortunately, instruction-driven image editing can significantly change the watermarked image at the semantic level, making \zj{current state-of-the-art watermarking methods ineffective}.

\zj{To remedy it,} we propose \Name, the first robust watermarking methodology against instruction-driven image editing. 
Specifically, \zj{we follow the classic structure of deep robust watermarking, consisting of the encoder, noise layer, and decoder.} 
To achieve robustness against semantic distortions, we introduce a novel Partial Instruction-driven Denoising Sampling Guidance (\ModuleName) module, which consists of a large variety of instruction injections and substantial modifications of images at different semantic levels. With \ModuleName, the encoder tends to embed the watermark into more robust and semantic-aware areas, which remains in existence even after severe image editing.
Experiments demonstrate that \Name can effectively extract the watermark from the edited image with a low bit error rate of nearly 2.6\% for 64-bit watermark messages. Meanwhile, it only induces a neglectable influence on the visual quality and editability of the original images. 
\textcolor{black}{Moreover, \Name holds general robustness against different sampling configurations and other popular image editing methods such as ControlNet-InstructPix2Pix, MagicBrush, Inpainting, and DDIM Inversion.} Codes and models are available at \href{https://github.com/hurunyi/Robust-Wide}{https://github.com/hurunyi/Robust-Wide}.
\end{abstract}

\section{Introduction}

Recently released Text-to-Image (T2I) diffusion models (e.g., GLIDE \cite{pmlr-v162-nichol22a}, DALL.E 2 \cite{ramesh2022hierarchical}, Imagen \cite{NEURIPS2022_ec795aea}, Stable Diffusion \cite{rombach2022high}) have pushed image generation capabilities to a new level. Trained on massive text-image pairs collected from the Internet, these models can generate high-quality photorealistic images based on given text prompts.
Instruction-driven image editing, a fantastic technique utilizing the power of T2I diffusion models, can edit the images on demand according to the instructions. Different models have been introduced to achieve this task. For instance, InstructPix2Pix \cite{Brooks_2023_CVPR} is a popular instruction-driven image editing model, which is fine-tuned from Stable Diffusion \cite{rombach2022high} on the dataset generated by GPT-3 \cite{NEURIPS2020_1457c0d6} and Prompt2Prompt \cite{hertz2022prompt}.
Afterwards, HIVE \cite{zhang2023hive}, MagicBrush \cite{zhang2023magicbrush}, and MGIE \cite{fu2023guiding} are proposed to improve InstructPix2Pix.

Despite the success of the instruction-driven image editing technique, these models could be misused by malicious users. First, attackers can exploit these models to modify normal images to create fake news, causing a crisis of trust in an individual or even a country. Typical examples include changing someone's face or expression, 
forging endorsements for commercial gain, or taking off the clothes to produce vulgar images.
Second, attackers can migrate the style based on a certain painting or make local modifications while keeping the basic composition of the painting unchanged to create new works, which shall be confirmed as plagiarism, infringing the IP rights of the work's owner.
By integrating the personalization technique (e.g., Textual Inversion \cite{gal2022image}), they may compose some concepts learned from other images in the edited image, which will undoubtedly lead to wider infringement on the concept's owners. 

To identify such misuse and trace malicious users, a common approach is watermarking. We can embed a secret watermark message into the original image, which can be extracted later for ownership verification. The embedded watermark must be robust enough against various  distortions. 
Traditional robust watermarking strategies
\cite{rahman2013dwt} mainly embed the watermark into a transformed domain to resist spatial distortions. To achieve robustness against more complex digital distortions, some deep watermarking methods are further proposed, e.g., HiDDeN \cite{Zhu_2018_ECCV} and MBRS \cite{jia2021mbrs}.
Additionally, researchers also introduced new methods to pursue robustness against physical distortions, including StegaStamp \cite{tancik2020stegastamp} and PIMoG \cite{fang2022pimog}.
Unfortunately, all the above solutions mainly target the pixel-level distortions, and fail to resist instruction-driven image editing, which induces significant distortions in the semantic level.

To remedy this issue, we propose \Name, the first \ul{robust} \ul{w}atermarking method for \ul{i}nstruction-\ul{d}riven image \ul{e}diting.
We are motivated by the popular encoder-noise layer-decoder framework in most deep watermarking methods \cite{Zhu_2018_ECCV,jia2021mbrs,tancik2020stegastamp,fang2022pimog}, which jointly achieve watermark embedding and extraction in an end-to-end way and leveragzhi lie the noise layer to simulate specific distortions to obtain the corresponding robustness. However, the main challenge in our task is how to simulate the distortions caused by instruction-driven image editing. To this end, we design a novel Partial Instruction-driven Denoising Sampling Guidance (\ModuleName) module in \Name. Briefly, \ModuleName allows the gradient of the last $k$ sampling steps to flow into the training pipeline, making the non-differentiable sampling process trainable. Additionally, it injects diverse instructions to guide distortions, forcing the encoder and decoder to focus on semantic areas for watermark embedding and extraction. 
%

We perform extensive experiments to demonstrate the robustness of \Name during the instruction-driven image editing process. It achieves a low Bit Error Rate (BER) of nearly 2.6\% for 64-bit watermark messages, while preserving the visual quality and editability of original images. Besides the robustness against semantic distortions, \Name acquires inherent robustness against pixel-level distortions such as JPEG and color shifting, which are unseen during training. 
It also holds general robustness against different sampling configurations and \zj{even different popular editing models such as ControlNet-InstructPix2Pix, MagicBrush, Inpainting \cite{inpaint} and DDIM Inversion \cite{mokady2023null}}. 

\textcolor{black}{
In summary, our contributions are as follows:
\begin{itemize}
    \item We point out the potential threat caused by the misuse of instruction-driven image editing and find current \zj{state-of-the-art} watermarking methods are ineffective in the emerging case. In other words, we unearth a novel robustness requirement for current image watermarking.
    \item We propose \Name, the first robust watermarking method for instruction-driven image editing. We introduce a novel Partial Instruction-driven Denoising Sampling Guidance (\ModuleName) module, \zj{which forces the watermark embedded in the semantic-level rather than pixel-level.}
    \item Experimental results demonstrate that \Name can resist instruction-driven image editing, conventional pixel-level distortions, and different sampling configurations. \zj{More importantly, the proposed method exhibits robustness against a variety of popular editing models.}
\end{itemize}
}

\section{Background}


\subsection{Diffusion Model}
Inspired by the non-equilibrium statistical physics, Diffusion Model (DM) \cite{pmlr-v37-sohl-dickstein15} destroys the structure in a data distribution through an iterative forward diffusion process, and learns a reverse diffusion process to restore data's structure. 
Denoising Diffusion Probabilistic Model (DDPM) \cite{NEURIPS2020_4c5bcfec} further improves the performance of DM
by training on a weighted variational bound with the folllowing objective:
\begin{eqnarray}
    L_{DM} = \mathbb{E}_{x, \epsilon\sim\mathcal{N}(0,1), t}\Big[\Vert \epsilon - \epsilon_\theta(x_t, t) \Vert_2^2\Big],
\end{eqnarray}
where $x$ is the input image, $\epsilon$ is the randomly sampled Gaussian noise, $t\in\{1,..., T\}$ is the uniformly sampled timestep, $x_t$ is the noisy version of $x$, and $\epsilon_\theta$ is the diffusion model trained to predict a denoised variant of $x_t$.

Many T2I diffusion models, e.g., GLIDE \cite{pmlr-v162-nichol22a}, DALL$\cdot$E 2 \cite{ramesh2022hierarchical} and Imagen \cite{NEURIPS2022_ec795aea} are based on DDPM.
They operate directly in the pixel space, which can consume a large amount of computational resources for both training and evaluation. To overcome these shortcomings,  Latent Diffusion Model (LDM) \cite{rombach2022high} is proposed to perform the noise and denoise process in the latent space of the pre-trained VAE:
\begin{eqnarray}
    L_{LDM} = \mathbb{E}_{\mathcal{E}(x), \epsilon\sim\mathcal{N}(0,1), t}\Big[\Vert \epsilon - \epsilon_\theta(z_t, t) \Vert_2^2\Big],
\end{eqnarray}
where $\mathcal{E}$ is the encoder of VAE and $z_t$ is the noisy latent variable.
Stable Diffusion is a popular T2I model based on LDM with great impact and outstanding performance.

\subsection{Instruction-driven Image Editing}
There are mainly five popular instruction-driven image editing models, which are all based on DMs. 
(1) InstructPix2Pix \cite{Brooks_2023_CVPR}
performs editing in a forward pass quickly and does not require any user-drawn mask, additional images, per-example fine-tuning, or inversion. It is trained in an end-to-end manner, with the dataset generated by GPT-3 \cite{NEURIPS2020_1457c0d6} and Prompt2Prompt \cite{hertz2022prompt}.
Each item in the dataset contains a source image, an editing instruction, and a subsequent edited image. During training, the image and instruction are regarded as conditions $c_I$ and $c_T$, respectively, while the edited image is the ground-truth output. Therefore,  the training object is as follows:
\begin{small}
\begin{eqnarray}
    L = \mathbb{E}_{\mathcal{E}(x),\mathcal{E}(c_I),c_T,\epsilon\sim\mathcal{N}(0,1),t}\Big[\Vert \epsilon - \epsilon_\theta(z_t, t, \mathcal{E}(c_I), c_T) \Vert_2^2\Big].
\end{eqnarray}
\end{small}
(2) ControlNet \cite{zhang2023adding} is a dedicated framework that amplifies the capability and controllability of pre-trained T2I diffusion models by integrating spatial conditioning controls.
Therefore, we can regard original images as spatial conditioning controls and train ControlNet on the dataset of InstructPix2Pix to realize instruction-driven image editing, which is called ControlNet-InstructPix2Pix \cite{zhang2023adding}.
(3) Afterwards, 
HIVE \cite{zhang2023hive} improves InstructPix2Pix by harnessing human feedback to tackle the misalignment between editing instructions and resulting edited images. \textcolor{black}{(4) MagicBrush \cite{zhang2023magicbrush} introduces the first large-scale and manually annotated dataset for instruction-guided real image editing and fine-tunes InstructPix2Pix on the dataset for better performance.} (5) MGIE \cite{fu2023guiding} uses multimodal large language models to derive expressive instructions and provide explicit guidance to further improve the editing performance while maintaining the efficiency.
Besides, \textcolor{black}{there are also some popular text-driven image editing methods such as Inpainting \cite{inpaint} and DDIM Inversion \cite{mokady2023null}. Inpainting edits an image by masking some regions of it and then regenerates the image according to the given text. DDIM Inversion \cite{mokady2023null} performs the reversed DDIM sampling process conditioned on the original text caption to convert the image to its partially noised version. Afterwards, the edited image is acquired by denoising the noised image given the edited text caption.}

\textcolor{black}{In this paper, we mainly target InstructPix2Pix, and assess the generalization of our method to ControlNet-InstructPix2Pix, MagicBrush, Inpainting and DDIM Inversion based on Stable Diffusion, since these models are open-sourced and publicly available. }


\subsection{Robust Watermarking}
Robust watermarking is widely used for IP protection and forensics. Traditional methods (\eg, DWT-DCT \cite{rahman2013dwt} and DWT-DCT-SVD \cite{rahman2013dwt}) embed a pre-defined watermark message into transformed domains to achieve simple robustness against different operations, e.g., affine transformation, scaling, etc. 
HiDDeN \cite{Zhu_2018_ECCV} is the first to leverage deep neural networks for watermark embedding and extraction. Importantly, it simulates JPEG compression into a differential module, inserted into the famous encoder-noise layer-decoder framework for robustness enhancement. RivaGAN \cite{zhang2019robust} applies a customized attention-based mechanism to embed diverse data and includes a separate adversarial network for optimizing robustness.
Afterwards, MBRS \cite{jia2021mbrs} utilizes a mixture of real JPEG, simulated JPEG, and noise-free images to get superior robustness performance against JPEG compression. 
\zj{Similarly, CIN \cite{ma2022towards} combines invertible and non-invertible mechanisms for robustness} to various digital distortions.
In addition to digital robustness, numerous works (LFM \cite{wengrowski2019light}, StegaStamp \cite{tancik2020stegastamp}, RIHOOP \cite{jia2020rihoop}, PIMoG \cite{fang2022pimog}) try to acquire robustness against physical operations such as screen-shooting, print-shooting, etc.
\textcolor{black}{Recently, SepMark \cite{10.1145/3581783.3612471} utilizes a unified framework for source tracing and Deepfake detection which also achieves robustness against face manipulations, a special form of image editing.}
However, none of them can resist instruction-driven image editing (\zj{see \Tref{tab:main results}}.)

In this paper, we compare our method with \zj{current state-of-the-art watermarking methods such as MBRS \cite{jia2021mbrs}, CIN \cite{ma2022towards}, PIMoG \cite{fang2022pimog}, and SepMark \cite{10.1145/3581783.3612471}.}
We also adopt DWT-DCT \cite{rahman2013dwt}, DWT-DCT-SVD \cite{rahman2013dwt}, and RivaGAN \cite{zhang2019robust} as the baseline methods, which are suggested by Stable Diffusion's official webpage \cite{sdiw}.

\begin{figure*}[t]
\centering
\includegraphics[width=\textwidth]{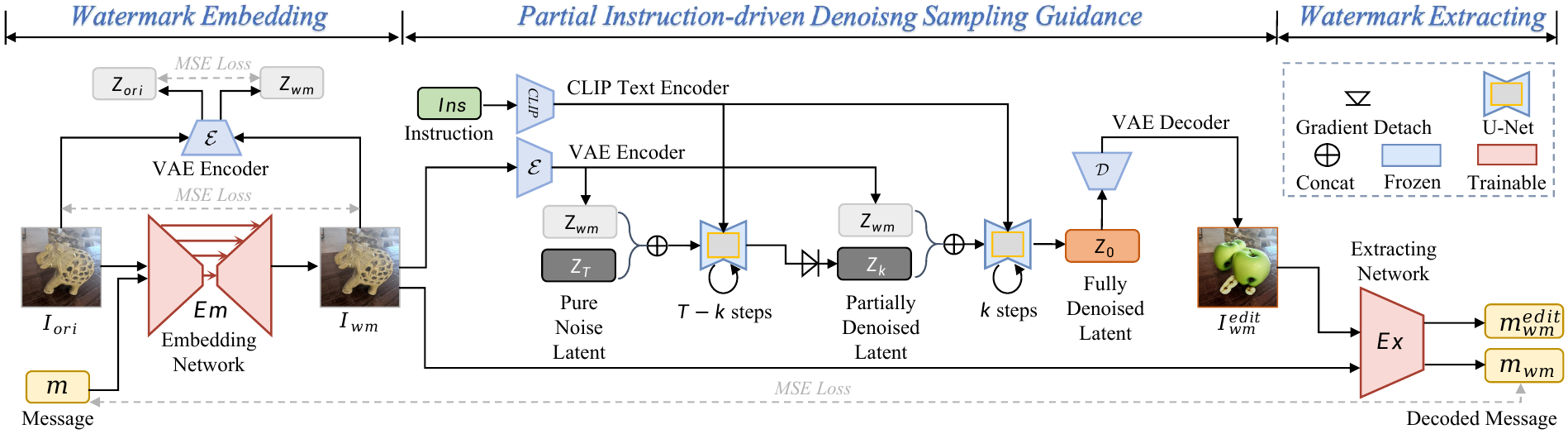}
\caption{The overall training pipeline of \Name.}
\label{fig:wm-pipeline}
\end{figure*}

\section{Methodology}

\subsection{Overview}

We introduce \Name, a novel methodology to embed robust watermarks into images, which can resist instruction-driven editing. Its overall training pipeline is shown in \Fref{fig:wm-pipeline}. 
An embedding network $E_m$ and extracting network $E_x$ are jointly trained to achieve watermark embedding and extraction, respectively. Furthermore, a novel Partial Instruction-driven Denoising Sampling Guidance (\ModuleName) module is integrated into the pipeline to enhance the watermark robustness against instruction-driven image editing. With the trained networks, we can use $E_m$ to embed a secret watermark message into a protected image, and release it to public. If a malicious user performs instruction-driven image editing over this watermarked image without authorization, we are able to detect this misuse by using $E_x$ to extract the watermark message from the edited image. Below we describe each step in detail.



\subsection{Watermark Embedding}
An embedding network $E_m$ is introduced to generate a watermarked image $I_{wm}$ from the original image $I_{ori}$, where the watermark is a random  $L$-bit message $m\in\{0,1\}^L$. 
Specifically, we adopt U-Net \cite{ronneberger2015u} as the structure of $E_m$. 
To match $m$ with the dimension of $I_{ori}$, $E_m$ converts the flattened version of $m$ in the shape of $1 \times \sqrt{L} \times \sqrt{L}$ to the shape of $C \times H \times W$ using some transposed convolutional layers, where $C$ is the predefined feature channel, $H$ and $W$ are the height and weight of $I_{ori}$. Then, we concatenate the reshaped message with $I_{ori}$. 
To preserve the editing capability of the image, $I_{wm}$ should be visually consistent with $I_{ori}$. We first adopt the $L_2$ distance between $I_{ori}$ and $I_{wm}$ in the pixel level, \ie, 
\begin{equation}
    L_{em_{1}} = L_2(I_{ori}, I_{wm}) = L_2(I_{ori}, E_m(I_{ori}, m)).
\label{eq:pixel-visual}
\end{equation}
Furthermore, we add another constraint between $I_{ori}$ and $I_{wm}$ in the feature space, represented by the encoder $\mathcal{E}$ of VAE in InstructPix2Pix, \ie,
\begin{eqnarray}
    L_{em_{2}}  = L_2(Z_{ori}, Z_{wm}) = L_2(\mathcal{E}(I_{ori}), \mathcal{E}(I_{wm})).
\end{eqnarray}

\subsection{Partial Instruction-driven Denoising Sampling Guidance}
\label{PIDSG}

Existing watermarking solutions mainly consider the robustness against pixel-level distortions. In contrast, instruction-driven image editing changes an image significantly at the semantic level, making these approaches ineffective.
We note that instruction-driven image editing involves the injection of a large number of different semantic instructions and varying degrees of image modifications at different semantic levels, which can be utilized to guide the robust and semantic-aware watermark embedding and extraction process. 
Thus, our intuitive idea is to incorporate the editing process into the end-to-end training framework. 
\textcolor{black}{However, one challenge is that during the denoising sampling process (\eg, in InstructPix2Pix), gradients are not allowed to flow directly. Introducing this process into training would result in the inability of gradients to propagate from the watermark decoder through the sampling process back to the watermark encoder. In other words, this would lead to a discontinuity in the computational graph, rendering the entire process non-differentiable.
While a straightforward approach might be to open gradients for the numerous denoising sampling steps, this would introduce a significant memory overhead.
To address this, we design the Partial Instruction-driven Denoising Sampling Guidance (\ModuleName) module, which selectively allows gradients to flow only in the last $k$ sampling steps. 
This design not only makes the entire method feasible but also enables the process to be differentiable and amenable to end-to-end optimization.}

As shown in the middle part of \Fref{fig:wm-pipeline}, InstructPix2Pix consists of VAE \cite{esser2021taming}, U-Net \cite{ronneberger2015u}, and CLIP text encoder \cite{radford2021learning}. We freeze all the parameters of these models.
%
%
During training, the encoder $\mathcal{E}$ of VAE converts $I_{wm}$ to its latent $Z_{wm}=\mathcal{E}(I_{wm})$. Then, $Z_{wm}$ is concatenated with the pure noise latent $Z_{T}$ and sent to U-Net to perform denoising sampling iterations.
Assuming the sampling process totally has $T$ steps, we truncate the gradient flow in the first $T-k$ steps to obtain the partially denoised latent $Z_k$. After that, we concatenate $Z_k$ with $Z_{wm}$ and perform the last $k$ sampling steps (dubbed gradient backward steps) to enable the gradient flow.
The CLIP text encoder processes the instruction $Ins$ and outputs the textual embedding to guide the whole sampling process. Finally, after $T$ sampling steps, the fully denoised latent $Z_0$ is produced and converted to the edited image $I^{edit}_{wm}$ by the decoder $\mathcal{D}$ of VAE.

\subsection{Watermark Extracting}
For the extracting network $E_x$, we leverage some residual blocks as its architecture. With the edited image $I^{edit}_{wm}$, $E_x$ aims to 
extract the message $m^{edit}_{wm}$ that is consistent to the original embedded message $m$, \ie,
\begin{eqnarray}
    L_{ex_{1}} = MSE(m, m^{edit}_{wm}) = MSE(m, E_x(I^{edit}_{wm})).
\end{eqnarray}
For effective forensic, we also require $E_x$ to be capable of extracting the embedded watermark message $m_{wm}$ from the watermarked image $I_{wm}$ before editing:
\begin{eqnarray}
    L_{ex_{2}}  = MSE(m, m_{wm}) = MSE(m, E_x(I_{wm})).
\end{eqnarray}
Interestingly, we observe that the training will not converge without $L_{ex_{2}}$. We explain that the extracting network $E_x$ cannot find the watermark area if only fed with edited images that are different from the original images at the semantic level. More results can be found in \Sref{ablation study}.

\subsection{Joint Training}
We jointly train $E_m$ and $E_x$ with the above-mentioned components: watermark embedding, \ModuleName, and watermark extraction. The total loss is formulated as follows:
\begin{eqnarray}
    L_{total} = L_{em_{1}} + \lambda_{1}L_{em_{2}} + \lambda_{2}L_{ex_{1}} + \lambda_{3} L_{ex_{2}},
\end{eqnarray}
where $\lambda_{1}=0.001$, $\lambda_{2}=0.1$, and $\lambda_{3}=1$ by default are the hyperparameters to balance each loss item. More analysis can be found in \Sref{ablation study}.

\section{Experiments}

\noindent\textbf{Datasets.}
To train the embedding and extracting networks, we adopt 20k image-instruction pairs from the dataset used in InstructPix2Pix.
We also select 1.2k additional samples that do not overlap with the above training data for evaluation by default.
Besides, we collect some real-world images from the Internet, which cover 6 types (\ie, person, animal, object, architecture, painting, and scenery) and each type has 5 images \textcolor{black}{which can be found in the supplementary material}. Then, we use InstructPix2Pix \cite{brooks2023instructpix2pix} \zj{by default} to edit these images based on 6 instructions 
and generate 8 images per instruction to finally obtain 1.44k edited images for testing.

\noindent\textbf{Implementation Details.}
We train all our models on a single A6000 GPU, with a learning rate of 1e-3, batch size of 2, and total steps of 20,000.  We use the AdamW optimizer with a cosine scheduler of 400 warm-up steps. Images used for training and evaluation are all 512$\times$512 by default. For the configurations of \ModuleName, we adopt the Euler sampler, with 20 inference steps, the text guidance scale $s_T$=10.0, and image guidance scale $s_I$=1.5.

\textcolor{black}{We choose seven baselines for comparisons, \ie, DWT-DCT \cite{rahman2013dwt}, DWT-DCT-SVD \cite{rahman2013dwt}, RivaGAN \cite{zhang2019robust}, MBRS \cite{jia2021mbrs}, CIN \cite{ma2022towards}, PIMoG \cite{fang2022pimog} and SepMark \cite{10.1145/3581783.3612471}. We directly use their official code for implementation. Noteably, the released model of MBRS and SepMark only support 256$\times$256 images with 256 bits and 128 bits separately and we used the Tracer of SepMark for watermark extraction after image editing. Since the released model of CIN only supports 128$\times$128 images and it is robust against the resize operation, so we resized the watermarked image to 256$\times$256 for image editing and then resized it back to 128$\times$128 for watermark extraction. PIMoG does not provide model weights, we trained the model ourselves using the official code.}

\noindent\textbf{Metrics.}
To evaluate the effectiveness of our method, we measure the Bit Error Rate (BER) between the extracted watermark $X$ and ground-truth watermark $Y$, \ie,
$BER(X, Y) = \frac{\sum^{L}_{i=1}(X_{i} \neq Y_{i})}{L}$,
where $X_{i}, Y_{i} \in \{0, 1\}$ and $L$ is the watermark length.
To assess the fidelity, we adopt PSNR and SSIM to measure the visual quality of watermarked images.
To verify how the watermarked image can preserve its original editability, we adopt the CLIP image similarity (CLIP-I) and CLIP Text-Image Direction Similarity (CLIP-T), which are also used in InstructPix2Pix \cite{Brooks_2023_CVPR}.

\begin{table*}[htb!]
\centering
\caption{Quantitative results compared with other methods. }
\footnotesize
\setlength\tabcolsep{5.5pt}
\scalebox{0.7}{
\begin{tabular}{l|cccccc}
\toprule
\multirow{2}{*}{\textbf{Method}} & \multirow{2}{*}{\textbf{Image Size}} & \textbf{Watermark} & \multicolumn{2}{c}{\textbf{BER (\%)} $\downarrow$} & \multirow{2}{*}{\textbf{PSNR}$\uparrow$} & \multirow{2}{*}{\textbf{SSIM}$\uparrow$} \\
& & \textbf{Length (bits)} & w/o Editing        & w/ Editing  & & \\
\midrule
DWT-DCT \cite{rahman2013dwt} & 512x512& 32 & 11.9351 & 49.2286 & 38.7123 & 0.9660 \\
DWT-DCT-SVD \cite{rahman2013dwt} & 512x512 & 32 & 0.0314 & 47.5680 & 38.6488 & 0.9726 \\
RivaGAN \cite{zhang2019robust} & 512x512 & 32 & 0.6276 & 40.5256 & 40.6132 & 0.9718 \\
MBRS \cite{jia2021mbrs} & 256x256 & 256 & 0.0000 & 46.7661 & 43.9780 & 0.9870 \\
CIN \cite{ma2022towards} & 128x128 & 30 & 0.0000 & 44.9888 & 40.3678 & 0.9763 \\
PIMoG \cite{fang2022pimog} & 256x256 & 64 & 0.0000 & 49.9635 & 35.3183 & 0.9212  \\
SepMark \cite{10.1145/3581783.3612471} & 256x256 & 128 & 0.0084 & 28.1460 & 36.4341 & 0.9194 \\
\Name  & 512x512 & 64 & 0.0000 & 2.6579 & 41.9142 & 0.9910 \\
\Name  & 512x512 & 256 & 0.0000 & 4.1867 & 39.1842 & 0.9844 \\
\bottomrule
\end{tabular}
}
\label{tab:main results}
\end{table*}

\subsection{Effectiveness Evaluation}

\begin{table}[htb!]
\centering
\caption{The importance of \ModuleName. $\dagger$ denotes the results on real-world images.}
\footnotesize
\setlength\tabcolsep{5.5pt}
\scalebox{0.6}{
\begin{tabular}{l|cccccc}
\toprule
\multirow{2}{*}{\textbf{Method}} & \multirow{2}{*}{\textbf{Image Size}} & \textbf{Watermark} & \multicolumn{2}{c}{\textbf{BER (\%)} $\downarrow$} & \multirow{2}{*}{\textbf{PSNR}$\uparrow$} & \multirow{2}{*}{\textbf{SSIM}$\uparrow$} \\
& & \textbf{Length (bits)} & w/o Editing        & w/ Editing  & & \\
\midrule
\Name (w/o \ModuleName) & 512x512 & 64 & 0.0000 & 50.1558 & 55.3710 & 0.9982 \\
\Name (w/ \ModuleName) & 512x512 & 64 & 0.0000 & 2.6579 & 41.9142 & 0.9910 \\
\midrule
\Name (w/o \ModuleName) $\dagger$ & 512x512 & 64 & 0.0000 & 51.0801 & 55.3715 & 0.9983 \\
\Name (w/ \ModuleName) $\dagger$ & 512x512 & 64 & 0.0000 & 2.6062 & 41.4038 & 0.9922 \\
\bottomrule
\end{tabular}
}
\label{tab:pidsg}
\end{table}

\begin{wraptable}{r}{0.5\linewidth}
\centering  
\caption{The integrity of \Name and the influence on the image editability.}  
\scalebox{0.4}{
\begin{tabular}{c|cccc}
\toprule
 Metrics & w/o Editing (\textbf{BER} (\%) $\downarrow$ )   & w/ Editing  (\textbf{BER} (\%) $\downarrow$ )   & \textbf{CLIP-I} $\uparrow$ & \textbf{CLIP-T} $\uparrow$  \\
\midrule
Original Images & 49.8403 & 48.8550  & 0.8402 & 0.2183 \\
Watermarked Images & 0.0000 & 2.6579 & 0.8430 & 0.2148  \\
\bottomrule 
\end{tabular}
}
\label{tab:integrity}
\end{wraptable}

Table \ref{tab:main results} compares the effectiveness of \Name with the baseline methods with different image sizes and watermark lengths. For \Name, we consider the implementation without and with \ModuleName. From this table, it is obvious that none of the baseline methods can resist instruction-driven image editing, with the BER of around 50\%. 
Comparably, \Name is effective with a low BER of 2.6579\%.
\zj{\Tref{tab:pidsg} also shows the effectiveness of \Name on 1.44k real-world samples. }
Importantly, the removal of \ModuleName leads to a complete failure, revealing the importance of \ModuleName.
Moreover, 
\Tref{tab:integrity} shows that \Name will not extract watermarks from original images with or without editing, guaranteeing forensic integrity.


\subsection{Fidelity Evaluation}

\begin{figure}
\centering
\includegraphics[width=0.7\textwidth]{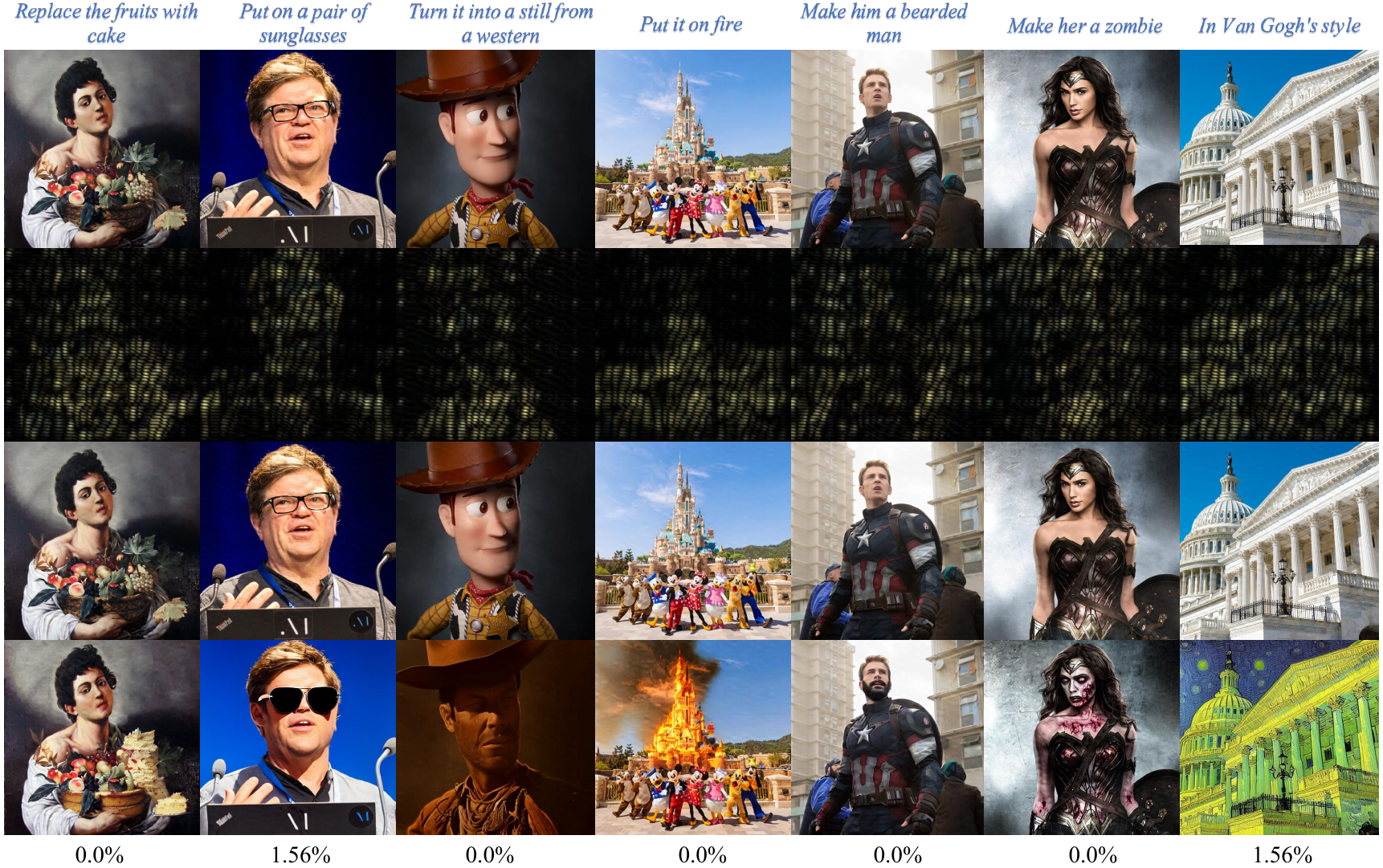}
\caption{Visual results for \Name. From top to bottom: instructions, original images, normalized residual images, watermarked images, edited images, and the corresponding BERs.}
\label{fig:main results}
\end{figure}

\Tref{tab:main results} also shows the PSNR and SSIM of different methods. We observe that \Name can achieve the comparable visual quality with other baseline methods.
\Tref{tab:integrity} compares the values of CLIP-I and CLIP-T for original and watermarked images. The slight difference indicates that \Name induces little influence on the editability. \Fref{fig:main results} shows some visual results using our \Name, which further confirms the fidelity (\textcolor{black}{more results are provided in the supplementary material}). Specifically, the normalized residual images are computed as $N(|I_{wm}-I_{ori}|)$, where $N(x) = (x-min(x))/(max(x)-min(x))$. From these images, it is evident that the watermark is predominantly embedded along the contours of the main subjects (such as people or objects) and in the background (secondary elements like buildings). We posit that these areas may be robust regions related to conceptual content and therefore \Name tends to embed watermarks into robust concept-aware areas.


\subsection{Robustness Evaluation}

\noindent\textbf{Pixel-level Distortions.}
\begin{table}[t]
\centering
\scriptsize
\caption{Robustness of \Name against pixel-level distortions.}
\scalebox{0.6}{
\begin{tabular}{c|c|c|c|c|c|c|c|c|c|c|c}
\hline
\textbf{BER(\%)} & \textbf{None} & \textbf{JPEG} & \textbf{Median Blur} & \textbf{Gaussian Blur} & \textbf{Gaussian Noise} & \textbf{Sharpness} & \textbf{Brightness} & \textbf{Contrast} & \textbf{Saturation} & \textbf{Hue} & \textbf{Noise+Denoise} \\
\hline
I & 2.6579 & 2.7256 & 2.4926 & 2.7074 & 3.0672 & 2.6150 & 12.2907 & 5.1336 & 3.0455 & 2.7591 & 8.6401 \\
II & 0.0000 & 0.0013 & 0.0000 & 0.0013 & 0.0716 & 0.0013 & 0.4733 & 0.9210 & 0.0000 & 0.0000 & 3.9071 \\
III & 2.6579 & 2.7934 & 2.9574 & 2.8489 & 6.0546 & 2.7773 & 9.4796 & 4.3346 & 3.0373 & 3.2829 & 9.3654 \\
\hline
\end{tabular}
}
\label{table:robustness normal distortions}
\end{table}
We apply different pixel-level distortions in three ways: (\uppercase\expandafter{\romannumeral1}) pre-processing watermarked images before editing; (\uppercase\expandafter{\romannumeral2}) post-processing watermarked images; (\uppercase\expandafter{\romannumeral3}) post-processing edited images based on watermarked images.
Table \ref{table:robustness normal distortions} shows the extraction error of \Name against different pixel-level distortion types in three scenarios. It is obvious that \Name demonstrates strong robustness against these distortions even if we do not involve them during training. 

Additionally, we also test \Name's robustness against DiffWA\cite{li2023diffwa}, a new watermark removal attack. This attack utilizes an image-to-image conditional diffusion model to add noise on the watermarked image and then restores the image while removing the embedded watermark. Due to the unavailability of DiffWA's source code, we utilize the open-source SDEdit\cite{meng2022sdedit} from Diffusers to carry out the process of adding noise and subsequently restoring the watermarked images. We configure the sampling parameters as follows: strength=0.2, prompt=Null, with all other sampling parameters as their default values. As shown in the ``Noise+Denoise" row of Table \ref{table:robustness normal distortions}, \Name is effective in all three situations with BER$<10\%$.



\noindent\textbf{Different Sampling Configurations.}
\begin{figure}[htb!]
\centering
\includegraphics[width=0.7\textwidth]{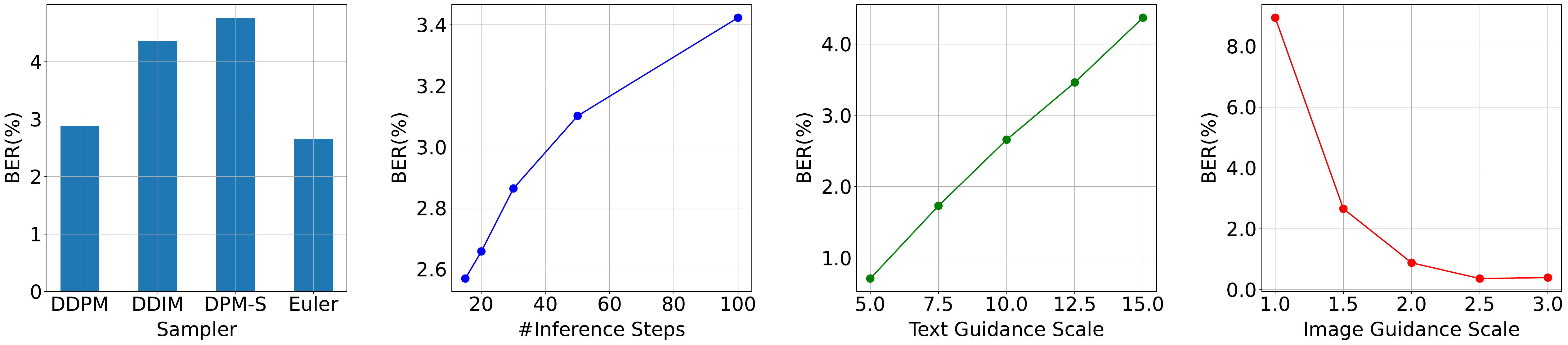}
\caption{Robustness of \Name against different diffusion sampling configurations.}
\label{fig:robustness again different sampling}
\end{figure}
We assess the robustness of \Name against different diffusion sampling configurations. 
\Fref{fig:robustness again different sampling} shows the corresponding results. We have the following observations. (1) \Name is generally effective for different samplers. (2)
As the total number of inference steps increases, the edited images become more fine-grained while the BER slightly increases (within a range of 1\%). We explain that more details 
lead to larger differences between the edited and original images. (3) 
A larger text guidance scale ($s_T$) and smaller image guidance scale ($s_I$) indicate more severe image editing. \Name achieves BER of below 5\% in all cases except when $s_I$ is 1, validating its general robustness in different settings.


\begin{figure}[htb!]
\centering
\includegraphics[width=0.8\textwidth]{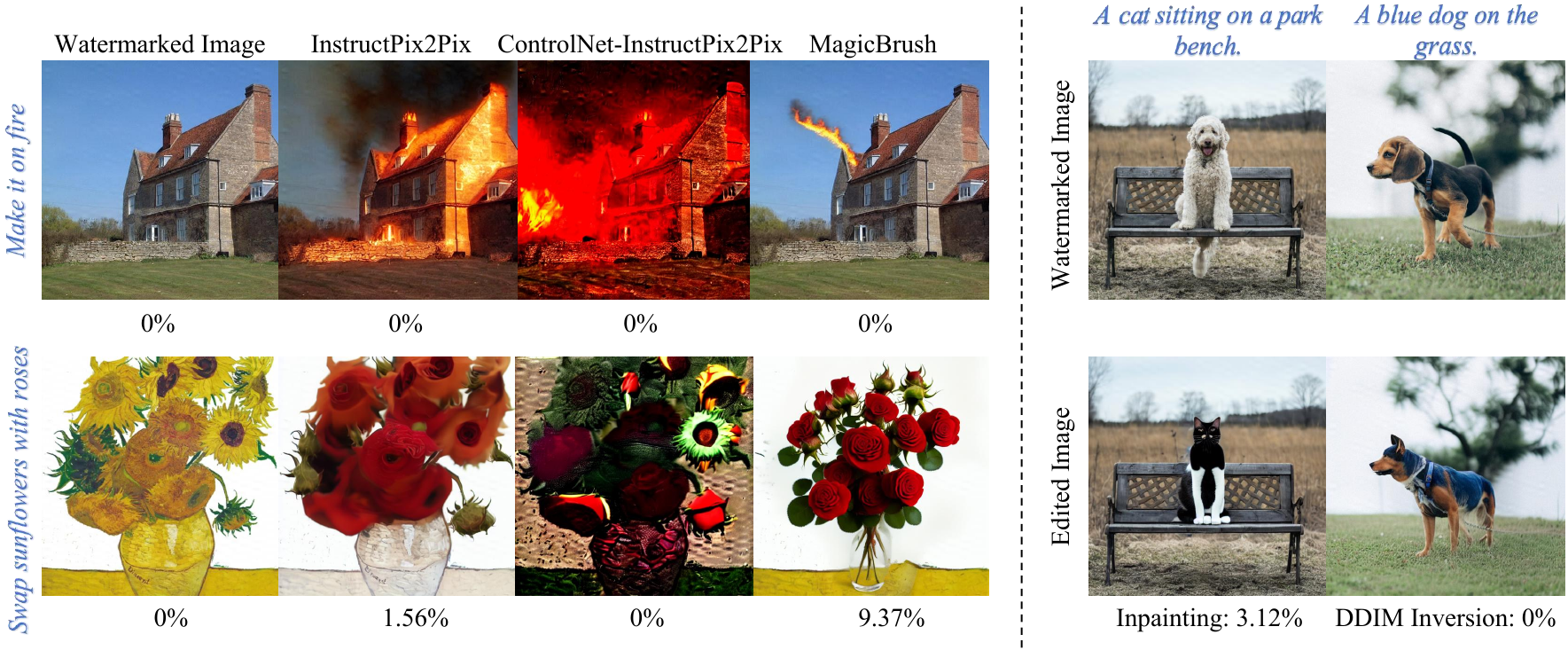}
\caption{\zj{General robustness against other editing methods such as  InstructPix2Pix, ControlNet-InstructPix2Pix, MagicBrush, Inpainting, and DDIM Inversion.}}
\label{fig:transfer study}
\end{figure}

\noindent{\textbf{Other Popular Image Editing Methods.}}
In addition to InstructPix2Pix, we further evaluate our robustness against its extension ControlNet-InstructPix2Pix \textcolor{black}{and MagicBrush} on the 1.2k-samples dataset. Experiments show that \Name can achieve an average BER of 0.96\% \textcolor{black}{on ControlNet-InstructPix2Pix and 9.34\% on MagicBrush}. \Fref{fig:transfer study} shows some visual examples (\textcolor{black}{more examples are provided in the supplementary material}).
For the instruction ``swap sunflowers with roses", MagicBrush makes the edited image more different from the original image, leading to a relatively higher but still acceptable BER.

\textcolor{black}{We also tested the robustness of Robust-Wide against Inpainting \cite{inpaint} and DDIM Inversion\cite{mokady2023null} based on Stable Diffusion Models. We found that, even though Robust-Wide has never seen these image editing methods during training, it still demonstrates effective resistance as shown in \Fref{fig:transfer study}.}

\noindent\textbf{Continual Editing.}
A user may utilize InstructPix2Pix to perform multiple edits on a single image. An image edited by one user could also be spread to another for further editing, and this process can repeat several rounds. Hence, we need to ensure our watermarks can resist continual editing. 
\begin{figure}[htb!]
\centering
\includegraphics[width=0.7\textwidth]{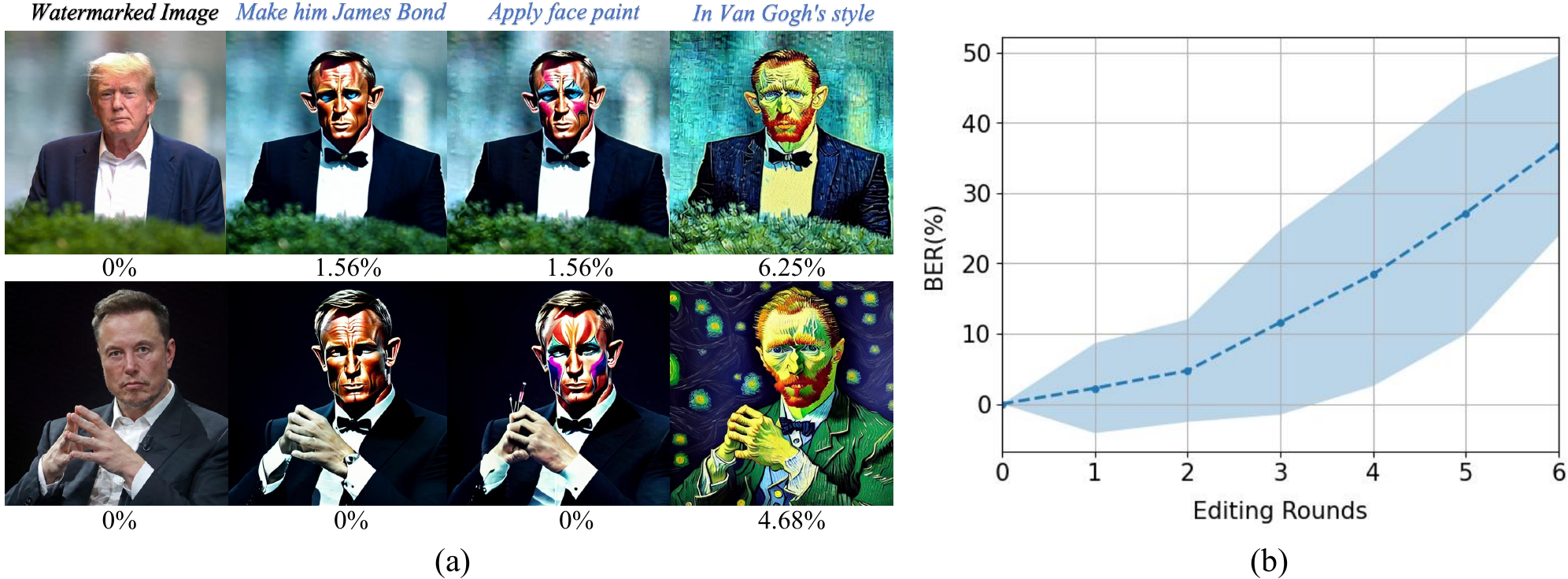}
\caption{The influence of continual editing. (a) Some visual examples under continual editing (from left to right). (b) BER increases with more editing rounds. 
This experiment is conducted on real-world images as mentioned above.}
\label{fig:multi edits}
\end{figure}
As shown in \Fref{fig:multi edits},
we observe that the watermark embedded into the original image can be accurately extracted even after 3 editing rounds, demonstrating the watermark's strong robustness against continual editing.

\subsection{More Analysis}

\begin{wrapfigure}{r}{0.6\linewidth}
\centering
\includegraphics[width=0.6\textwidth]{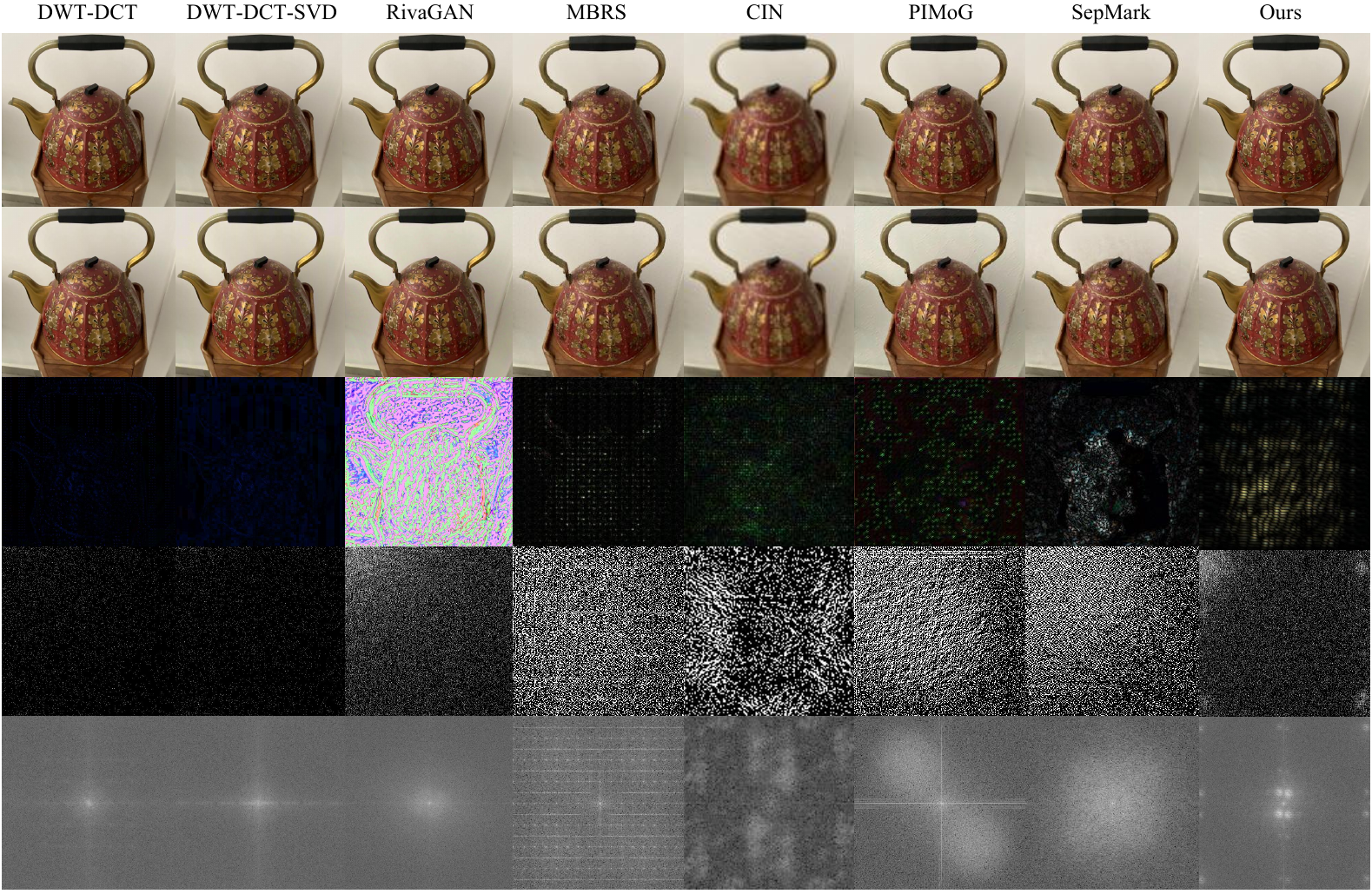}
\caption{The embedding mode of different methods. From top to bottom:  original images, watermarked images, normalized residual images, DCT and DFT of residual images, respectively}
\label{fig:wm pattern}
\end{wrapfigure}

\noindent\textbf{Embedding mode of \Name.}
\Fref{fig:wm pattern} displays the normalized residual image between the watermarked and original images, using different watermarking methods. We can see that the watermarks embedded using DWT-DCT and DWT-DCT-SVD are quite faint and imperceptible, indicating their limited robustness. MBRS tends to concentrate primarily along the object's edge contours, making it vulnerable to changes in the image background. In contrast, RivaGAN and our \Name embed watermarks in both the edges and backgrounds, resulting in enhanced robustness.
Furthermore, we use Discrete Cosine Transform (DCT) and Discrete Fourier Transform (DFT) to visualize the residual images in the frequency domain.
\zj{From the residual images, we can observe that DWT-DCT and DWT-DCT-SVD introduce input-agnostic modification while other methods introduce input-aware modification. From the last two rows of \Fref{fig:wm pattern}, we find that DWT-DCT, DWT-DCT-SVD, and RivaGAN tend to mainly modify the low-frequency area while MBRS, CIN, PIMoG, and SpeMark prefer to embed the watermark into both low-frequency and high-frequency areas. Differently, \Name mainly focuses on embedding watermarks in infra-low frequency areas or mid-low frequency areas, potentially making it more robust against semantically and conceptually related image modifications. }


\begin{center}
  \begin{minipage}{0.3\textwidth}
    \centering
        \includegraphics[width=1.0\textwidth]{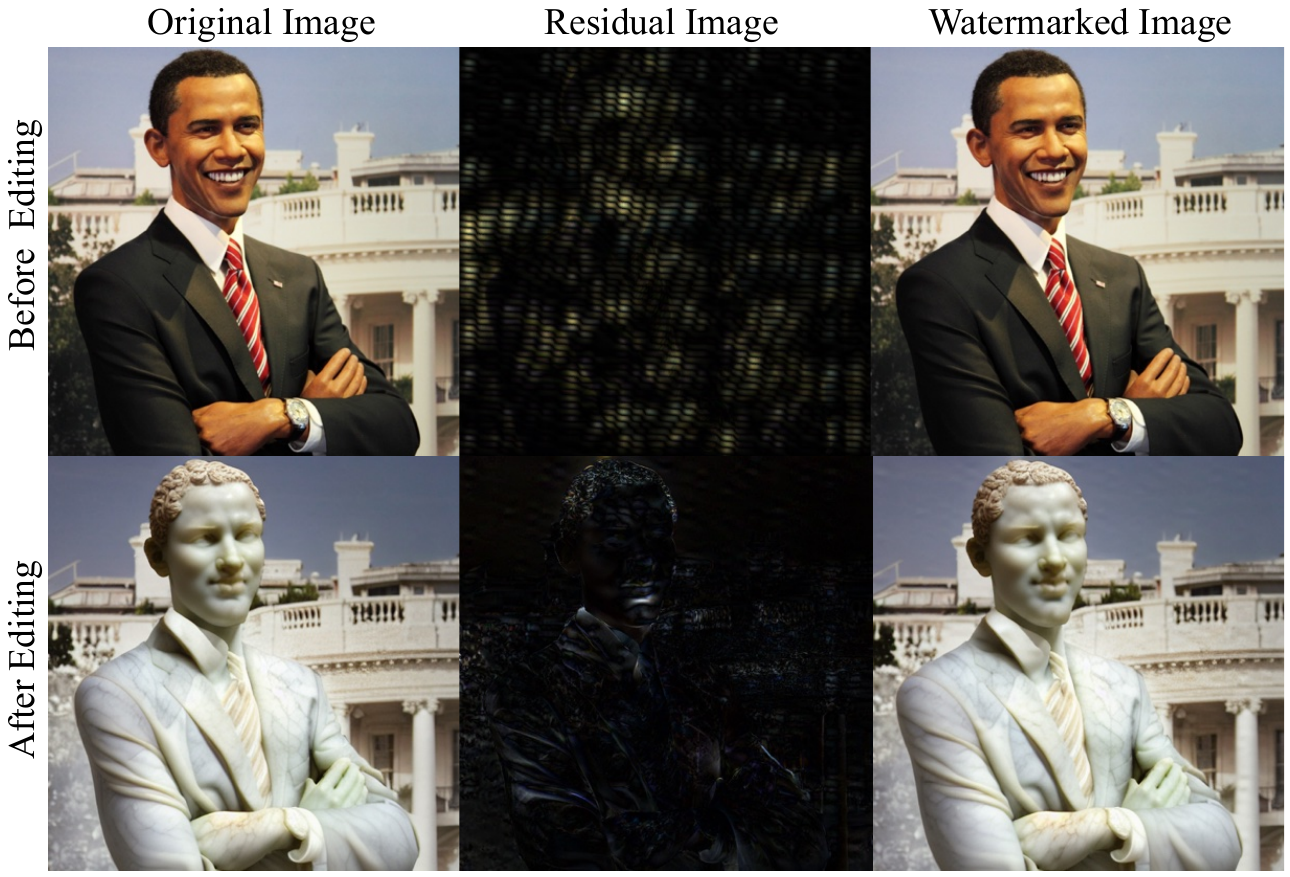}
        \captionof{figure}{Example of our extracting mode. }
        \label{fig:edit wm pattern}
  \end{minipage}\hfill
  \begin{minipage}{0.65\textwidth}
  \centering
    \includegraphics[width=1.0\textwidth]{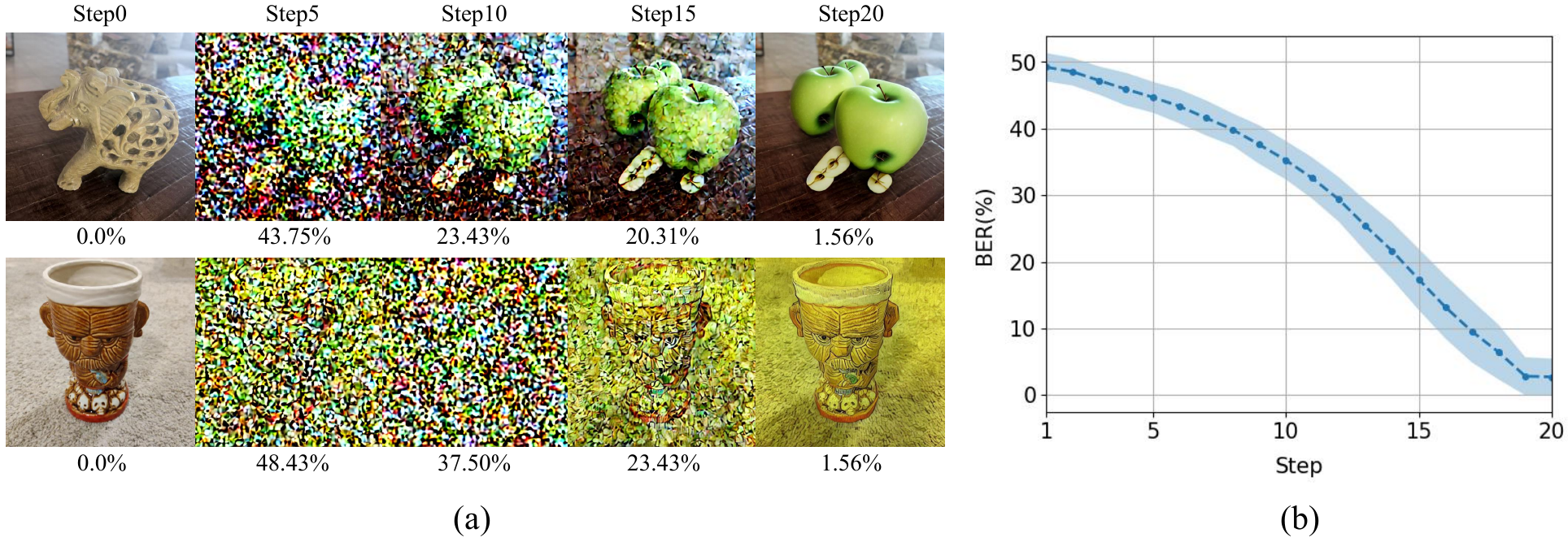}
    \captionof{figure}{The influence of the number of sampling steps.
(a) Some visual examples at different steps and their corresponding BER. (b) BER decreases with more steps.}
\label{fig:step ber}
  \end{minipage}
\end{center}


\noindent\textbf{Extracting mode of \Name.}
We use the original image $I_{ori}$ and watermarked image $I_{wm}$ to generate the corresponding edited images $I_{ori}^{edit}$ and $I_{wm}^{edit}$. Here, we control all potential random factors (\eg, generation random seed) to guarantee that the difference between $I_{ori}^{edit}$ and $I_{wm}^{edit}$ is solely caused by instruction-driven image editing.
As shown in \Fref{fig:edit wm pattern}, the watermark embedded into the original image also exists after editing, which preserves as an outline of the character. We hypothesize that the extractor mainly focuses on such robust concept-aware areas. \textcolor{black}{More examples are provided in the supplementary material.}


\noindent\textbf{Relationship between the number of sampling steps and extraction ability.}
We test the BER of generated images at different sampling steps.
\Fref{fig:step ber} (a) indicates that the model focues on generating contours and layout at the first few sampling steps, and then optimizing image details at later steps. As shown in \Fref{fig:step ber} (b), the BER decreases with more sampling steps. 

\noindent\textbf{Relationship between editing strength and extraction ability.}
\begin{wrapfigure}{r}{0.4\textwidth}
\centering
\includegraphics[width=0.4\textwidth]{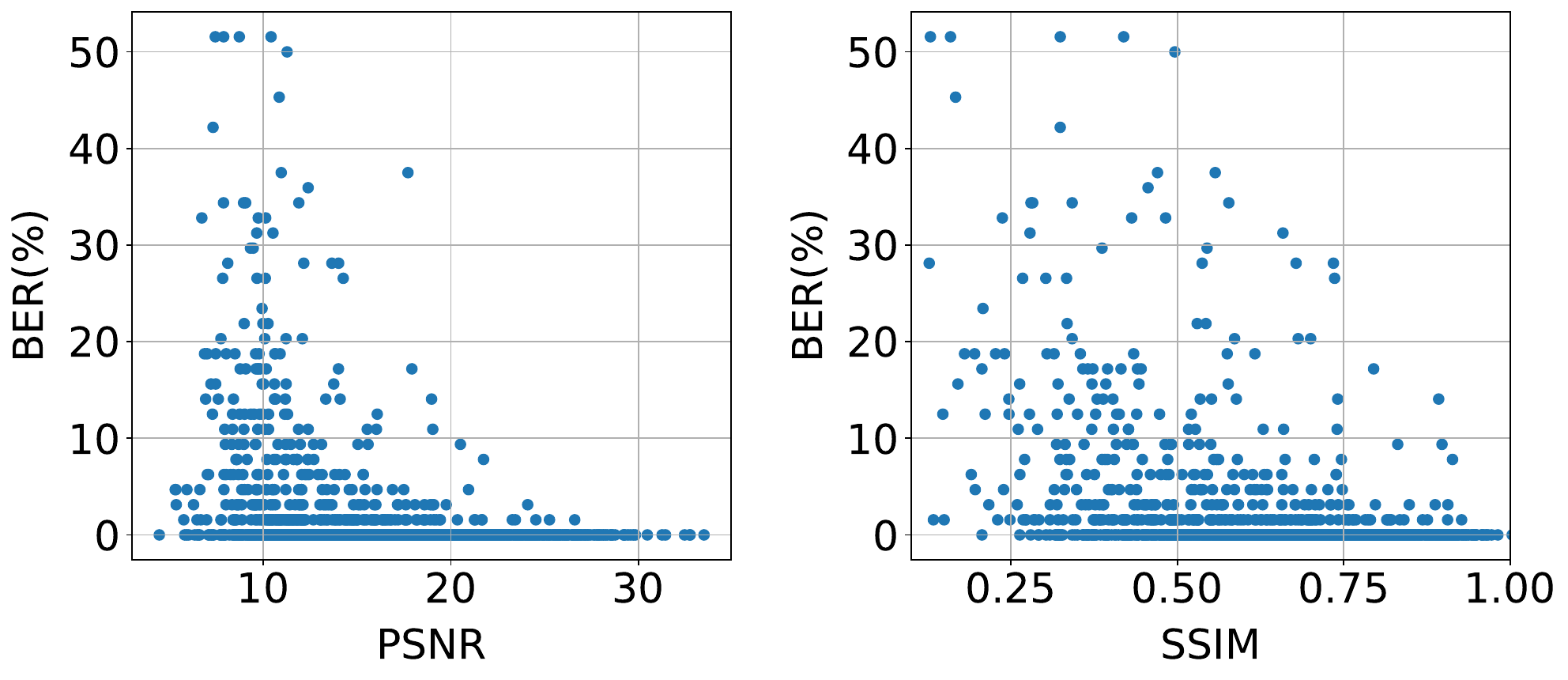}
\caption{The relationship between image editing strength and extraction ability.}
\label{fig:psnr ssim and ber}
\end{wrapfigure}
We explore the impact of the image editing strength (quantified by PSNR and SSIM computed between $I_{wm}$ and $I^{edit}_{wm}$) on the extraction BER. As shown in Figure \ref{fig:psnr ssim and ber}, points with higher BER are mainly at the area where PSNR and SSIM are low, \ie, the editing strength is large. This is intuitive as the greater the change is, the more difficult it will be for the watermark extraction. When the editing strength on the original image is significant, the resulting edited image can be considered as a form of re-creation. In such cases, the risk of copyright infringement is relatively low, and the ineffective extraction may be deemed acceptable.

\subsection{Ablation Study\label{ablation study}}

\noindent\textbf{Importance of $L_{ex_2}$.}
With $L_{ex_2}$, the embedding network and extracting network tend to find desirable watermarking areas at first and then search more robust areas to resist instruction-driven image editing. 
Without $L_{ex_2}$, it is challenging to achieve watermark embedding and extraction only with edited images. Table \ref{table:loss before sample} compares the performance under these two settings. The embedded watermark cannot be effectively extracted without $L_{ex_2}$, both before and after editing, with the BER of around 50\%.
Hence, $L_{ex_2}$ is essential for the overall effectiveness of \Name.

\begin{wraptable}{r}{0.3\linewidth}
                \centering
                \scriptsize
                \caption{The importance of $L_{ex_2}$. The gray cell denotes the default setting.}                
                \scalebox{0.6}{
                \begin{tabular}{cc|c|c}
                \hline
                \multicolumn{2}{c|}{Metrics}    & w/o \textbf{$L_{ex_2}$}  & \cellcolor[gray]{0.9}w/ \textbf{$L_{ex_2}$}  \\ \hline
                \multirow{2}{*}{\textbf{BER(\%)$\downarrow$}} & w/o editing & 50.1098 & 0.0000  \\
                                     & w/ editing  & 50.1219 & 2.6579  \\ \hline
                \multicolumn{2}{c|}{\textbf{PSNR$\uparrow$}}         & 68.7220 & 41.9142 \\ \hline
                \multicolumn{2}{c|}{\textbf{SSIM$\uparrow$}}         & 0.9999  & 0.9910  \\ \hline
                \end{tabular}
                }
                \centering
                \label{table:loss before sample}
\end{wraptable}

\noindent\textbf{Impact of hyper-parameters $\lambda_1$ and $\lambda_2$.}
\Tref{table:loss weight} shows the watermark performance with different $\lambda_1$ and $\lambda_1$ values. We observe that a larger $\lambda_1$ can help improve the visual quality of watermarked images but causes a decrease in the watermark extraction rate. 
On the other hand, a higher $\lambda_2$ leads to lower BER,  which showcases the trade-off between extraction ability and visual quality.


\begin{minipage}{0.45\textwidth}
\scriptsize
\captionof{table}{The impact of different $\lambda_1$ and $\lambda_2$.} 
\scalebox{0.6}{
\begin{tabular}{c|cccc|ccc}
\hline
\multirow{2}{*}{Metrics}   & \multicolumn{4}{c|}{\textbf{$\lambda_1$}}                                                           & \multicolumn{3}{c}{\textbf{$\lambda_2$}}                                \\  
                    & \multicolumn{1}{c}{0} & \multicolumn{1}{c}{\cellcolor[gray]{0.9}0.001} & \multicolumn{1}{c}{0.01} & 0.1     & \multicolumn{1}{c}{0.01} & \cellcolor[gray]{0.9}0.1 & 1       \\ \hline
\textbf{BER(\%)$\downarrow$} & 1.7391                 & 2.6579                      & 6.0690                    & 50.0364 & 11.1796                   & 2.6579                   & 0.9674  \\ \hline
\textbf{PSNR$\uparrow$}      & 39.6859                & 41.9142                     & 41.7604                   & 56.9096 & 45.8701                   & 41.9142                  & 35.2577 \\ \hline
\textbf{SSIM$\uparrow$}      & 0.9750                 & 0.9910                      & 0.9938                    & 0.9986  & 0.9929                    & 0.9910                   & 0.9616  \\ \hline
\end{tabular}}
\label{table:loss weight}
\end{minipage}\hfill
\begin{minipage}{0.45\textwidth}
\scriptsize
\captionof{table}{The influence of $k$ and the bits length.}
\scalebox{0.6}{
\begin{tabular}{c|ccc|cccc}
\hline
\multirow{2}{*}{Metrics}   & \multicolumn{3}{c|}{\textbf{$k$}} & \multicolumn{4}{c}{\textbf{Bits Length}} \\  
                    & 1               & 2               & \cellcolor[gray]{0.9}3              & 16        & \cellcolor[gray]{0.9}64        & 256      & 1024     \\ \hline
\textbf{BER(\%)$\downarrow$} & 4.0520          & 2.9166          & 2.6579         & 2.2812    & 2.6579    & 4.1867   & 6.3036   \\ \hline
\textbf{PSNR$\uparrow$}      & 44.3330         & 42.1386         & 41.9142        & 40.8327   & 41.9142   & 39.1842  & 36.5956  \\ \hline
\textbf{SSIM$\uparrow$ }     & 0.9938          & 0.9919          & 0.9910         & 0.9853    & 0.9910    & 0.9844   & 0.9732   \\ \hline
\end{tabular}}
\label{tab:step_bit}
\end{minipage}

\noindent\textbf{Influence of the number of gradient backward steps $k$.}
Table \ref{tab:step_bit} shows the watermark performance with different numbers of gradient backward steps k. 
With more steps, the watermark message is easier to extract while the visual quality of the watermarked image slightly reduces.
Due to the GPU memory constraints, the maximum number of gradient backward steps we could set is 3, which is sufficient to obtain acceptable performance.

\noindent\textbf{Influence of different watermark bits length.}
Table \ref{tab:step_bit} reports the evaluation results with different lengths of watermark messages. A longer watermark message results in higher BER and lower visual quality. 
In practice, the user can customarily select the watermark lengths to balance such trade-off.
\textcolor{black}{In \Tref{tab:step_bit}, we can observe that the visual quality is lower when the watermark length is 16 compared to when the watermark length is 64. Indeed, the integration of watermark bit with the image involves shape transformations through convolutional or deconvolutional layers. With 16 bits, our available GPU memory posed restrictions on employing convolutional or deconvolutional layers for shape transformations. Consequently, we opted for a non-parametric interpolation method to handle the integration, causing degradation in visual quality.}




\section{Conclusion}
In this paper, we propose \Name, the first robust watermarking method against instruction-driven image editing. Our core idea is to integrate a novel module called \ModuleName into the encoder-noise layer-decoder training framework, which forces watermarks embedded in the semantic level. Experiments demonstrate that \Name is robust against different image editing methods while maintaining high visual quality and editability. 
Furthermore, our in-depth analysis on the embedding and extracting modes of \Name is expected to shed light on the design of watermarking against other semantic distortions.  

\section*{Acknowledgements}
This study is supported under the RIE2020 Industry Alignment Fund – Industry Collaboration Projects (IAF-ICP) Funding Initiative, as well as cash and in-kind contribution from the industry partner(s), and Singapore Ministry of Education (MOE) AcRF Tier 2 MOE-T2EP20121-0006.

%
%
\bibliographystyle{splncs04}
\bibliography{main}
\end{document}


\title{\Name: Robust Watermarking against Instruction-driven Image Editing}

\titlerunning{Robust Watermarking against Instruction-driven Image Editing}

\author{Runyi Hu\inst{1,2}\orcidlink{0009-0001-6974-2542} \and
Jie Zhang\inst{1}\thanks{The corresponding author}\orcidlink{0000-0002-4230-1077} \and
Ting Xu\inst{3} \and
Jiwei Li\inst{4} \and
Tianwei Zhang\inst{1}\orcidlink{0000-0001-6595-6650}
}


\institute{Nanyang Technological University \\
\email{runyi\_hu@163.com} \\
\email{\{jie\_zhang, tianwei.zhang\}@ntu.edu.sg} \and
S-Lab, NTU \and
National University of Singapore \\
\email{xuting@nus.edu.sg} \and
Zhejiang University \\
\email{jiwei\_li@zju.edu.cn}
}

\maketitle

\section{Real-World Images Dataset}
Here, we present a comprehensive overview of the construction process for the real-world images dataset. To encompass a diverse range of image modifications reflecting real-world scenarios, we opted for six categories of images from the Internet: person, animal, object, architecture, painting, and scenery. For each type, we chose five images and designed six hand-crafted editing instructions, respectively. More details are provided in the following part.

\subsection{Real-World Images of Different Types}
\begin{figure*}[htbp]
\centering
\includegraphics[width=\textwidth]{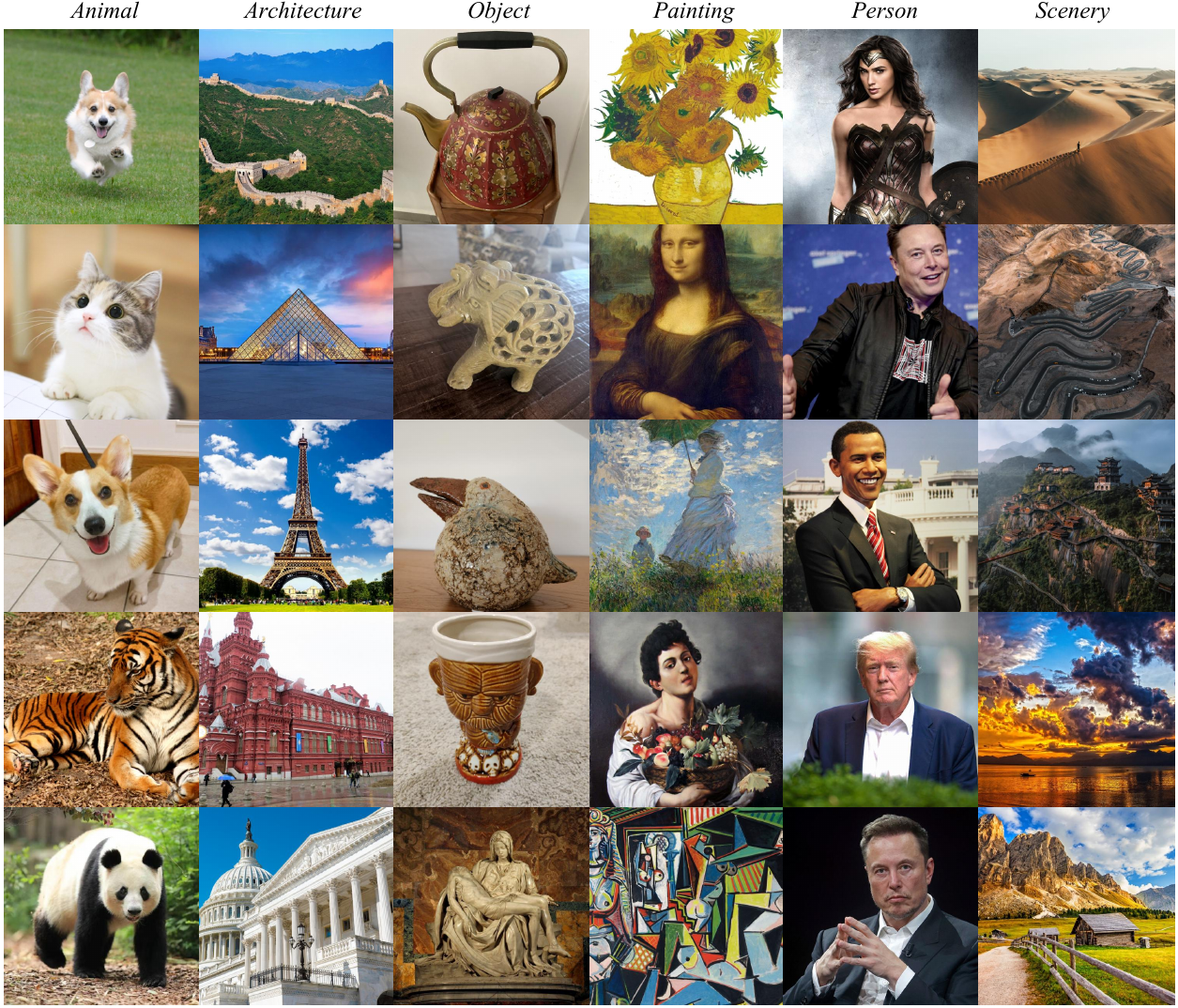}
\caption{The real-world images of six types.}
\label{fig:real images}
\end{figure*}

We show real-world images of different types in \Fref{fig:real images}.

\subsection{Hand-crafted Editing Instructions}
We show the designed editing instructions for each type.

\textbf{Animal}

\textit{``Move it to the desert''}

\textit{``Put on a pair of sunglasses''}

\textit{``Make it a cartoon''}

\textit{``Make it in Van Gogh's artistic style''}

\textit{``Make it Picasso style''}

\textit{``Make it Minecraft''}

\textbf{Architecture}

\textit{``Make it an Egyptian sculpture''}

\textit{``Turn it into ruins''}

\textit{``Put it on fire''}

\textit{``Make it Picasso style''}

\textit{``Make it in Van Gogh's artistic style''}

\textit{``Make it Minecraft''}

\textbf{Object}

\textit{``Replace it with an apple''}

\textit{``Change it with a rose''}

\textit{``Move it on the moon''}

\textit{``Make it a Modigliani painting''}

\textit{``Turn this into the space age''}

\textit{``Make it Minecraft''}

\textbf{Painting}

\textit{``Make it a Modigliani painting''}

\textit{``Make it a Miro painting''}

\textit{``Make it Picasso style''}

\textit{``Make it a Van Gogh's painting''}

\textit{``Make it Minecraft''}

\textit{``Make it a cartoon''}

\textbf{Person}

\textit{``Apply face paint''}

\textit{``Make it a zombie''}

\textit{``Take off the clothes''}

\textit{``Make it a marble roman sculpture''}

\textit{``It should look 100 years old''}

\textit{``Make it Minecraft''}

\textbf{Scenery}

\textit{``Add a beautiful sunset''}

\textit{``Cover it with snow''}

\textit{``Make it a cyberpunk painting''}

\textit{``Make it a Van Gogh's painting''}

\textit{``Make it Minecraft''}

\textit{``Make it a pencil drawing''}

\section{More Visual Examples}
\subsection{Main Results.} 
\begin{figure*}[htbp]
  \centering
  \begin{subfigure}[b]{\textwidth}
  \centering
    \includegraphics[width=0.9\textwidth]{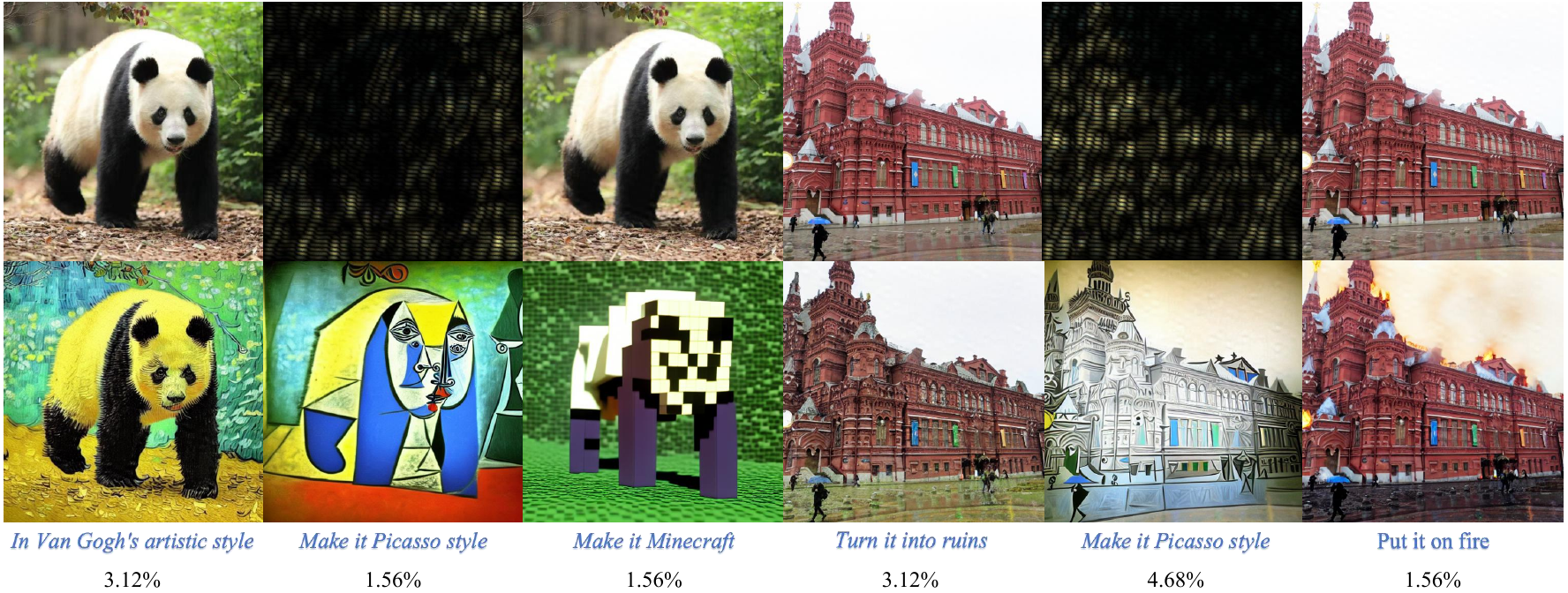}
    \label{fig:sub1}
  \end{subfigure}
  \begin{subfigure}[b]{\textwidth}
  \centering
    \includegraphics[width=0.9\textwidth]{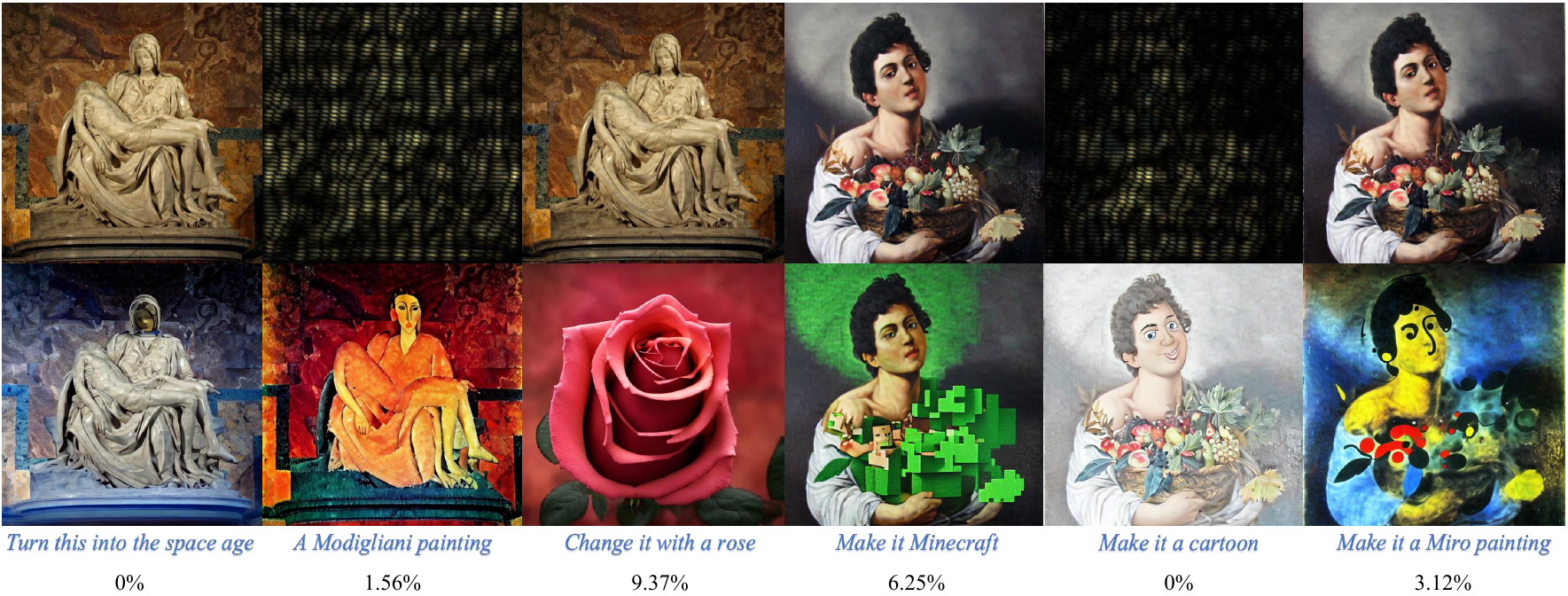}
    \label{fig:sub2}
  \end{subfigure}
  \begin{subfigure}[b]{\textwidth}
  \centering
    \includegraphics[width=0.9\textwidth]{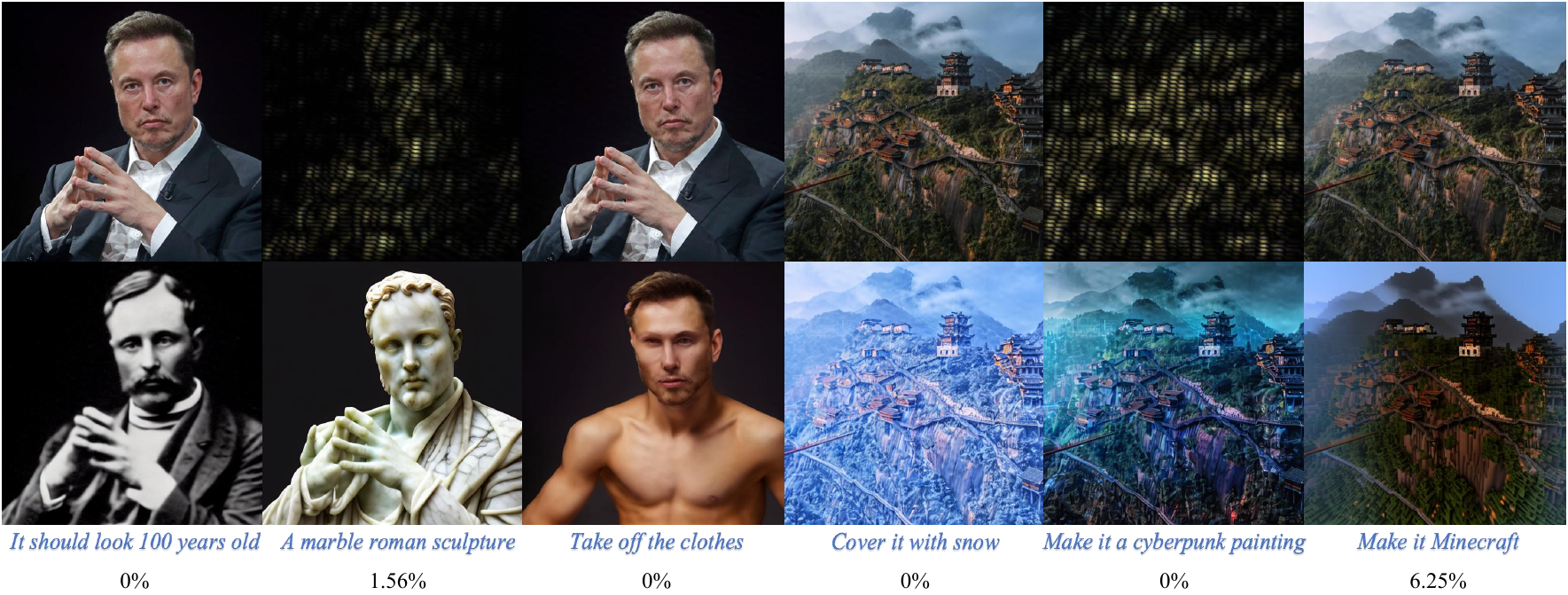}
    \label{fig:sub3}
  \end{subfigure}
  
  \caption{More visual examples from real-world images of six types, namely, animal, architecture, object, painting, person, and scenery. For each, the first row shows Original, Normalized Residual, and Watermarked Images and the second row shows Edited Images with the corresponding instructions, while BER is noted below.}
  \label{fig:main results}
\end{figure*}

See \Fref{fig:main results}.

\captionsetup{
    justification=raggedright,
    singlelinecheck=false
}

\captionsetup{
    justification=centering,
    singlelinecheck=false
}

\subsection{Robustness Against Different Instruction-driven Image Editing Methods.} 
\begin{figure*}[htbp]
  \centering
  \begin{subfigure}[b]{\textwidth}
  \centering
    \includegraphics[width=0.9\textwidth]{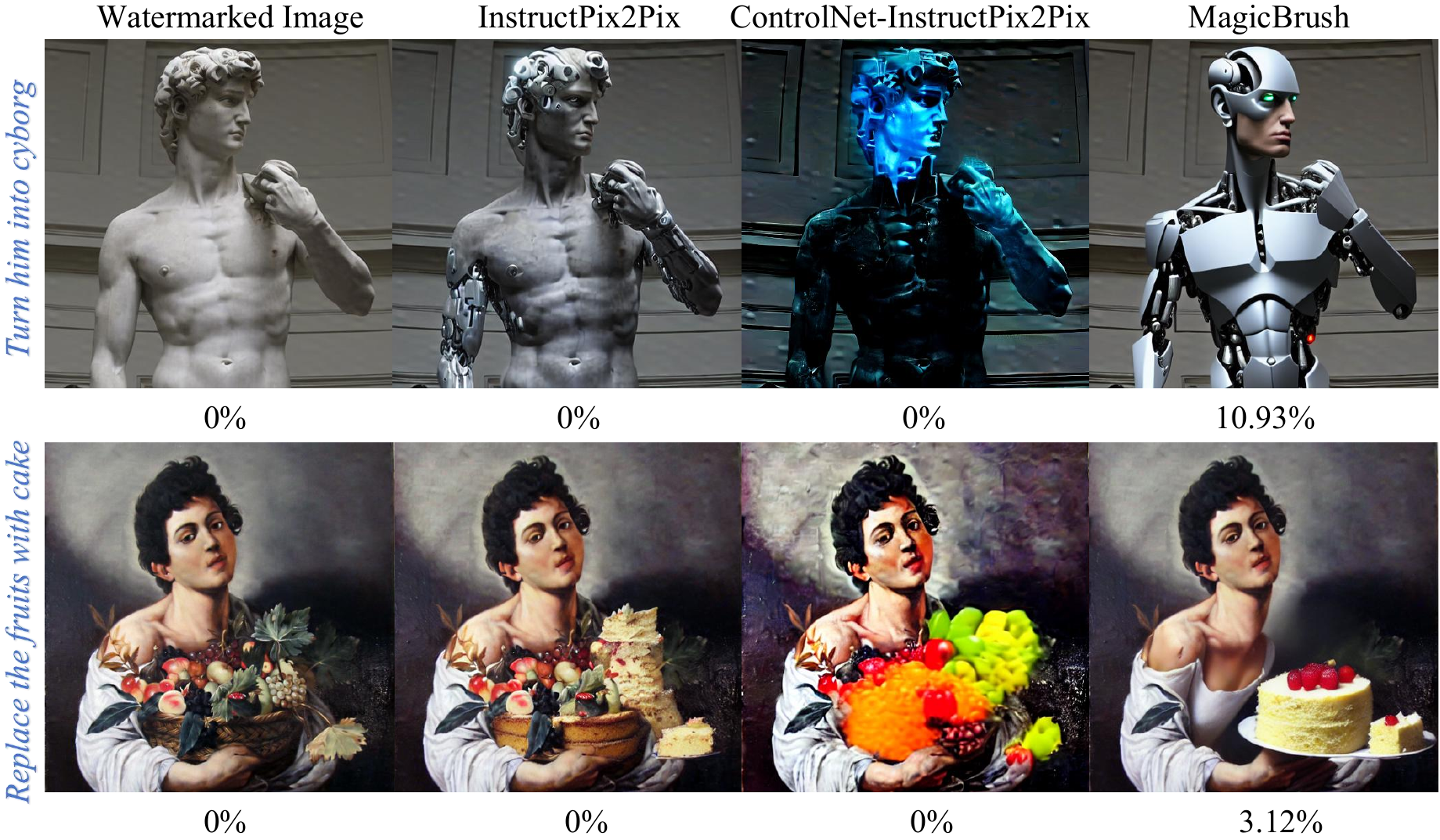}
    \label{fig:sub1}
  \end{subfigure}


  \begin{subfigure}[b]{\textwidth}
  \centering
    \includegraphics[width=0.9\textwidth]{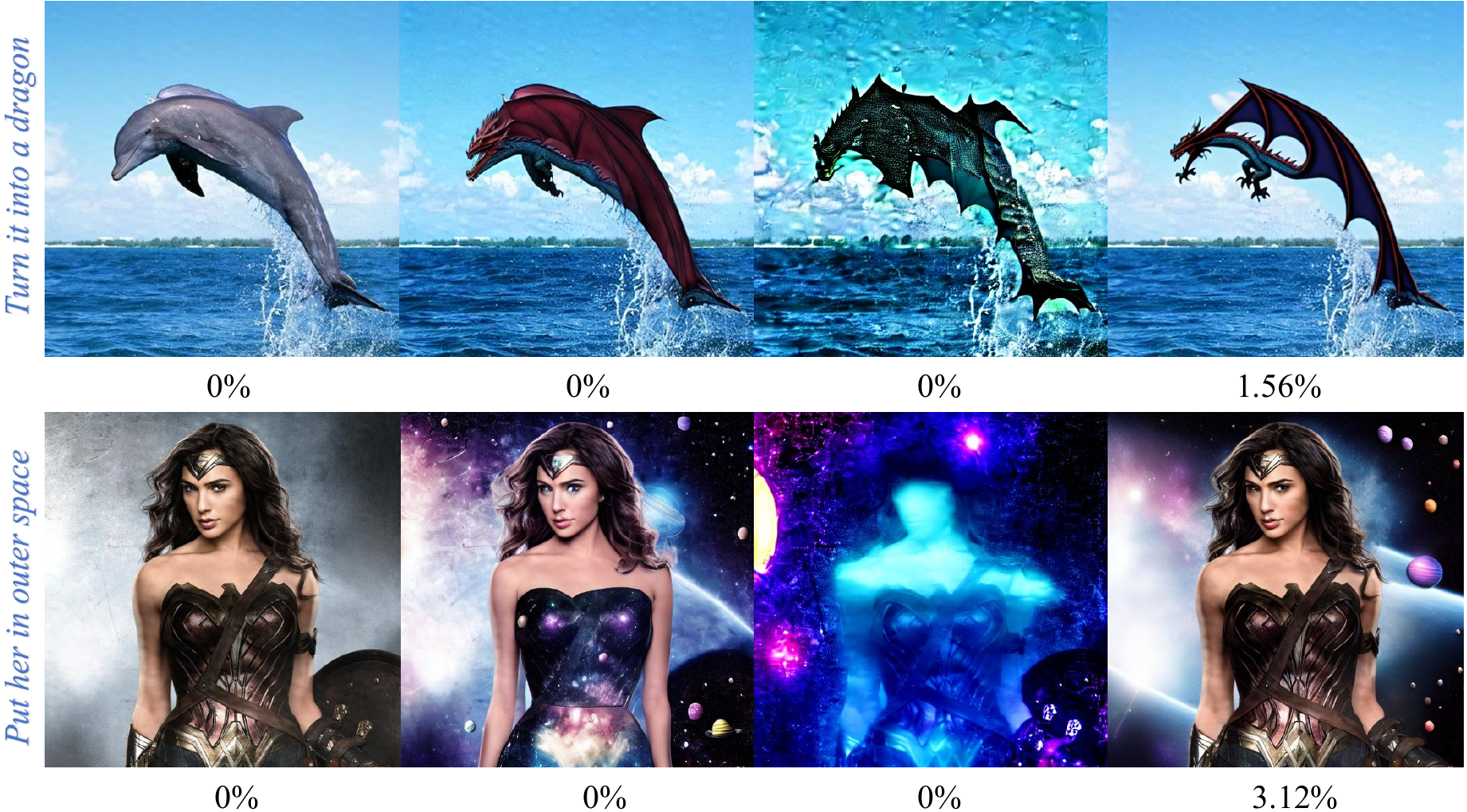}
    \label{fig:sub2}
  \end{subfigure}

    
  \begin{subfigure}[b]{\textwidth}
  \centering
    \includegraphics[width=0.9\textwidth]{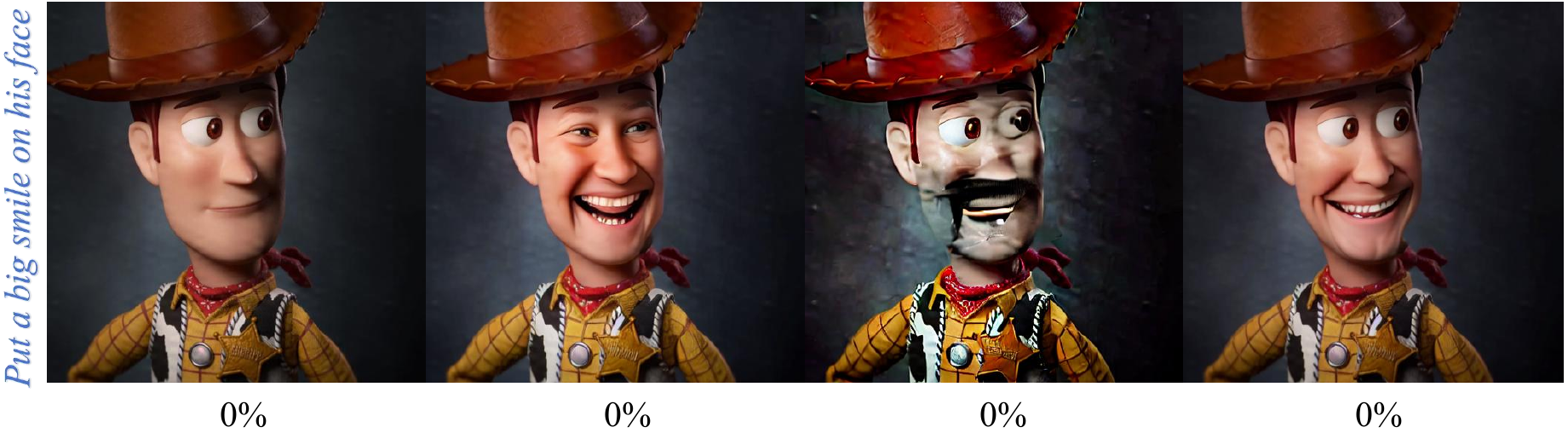}
    \label{fig:sub3}
  \end{subfigure}
  
  \caption{More visual examples of our robustness against different instruction-driven image editing methods.}
  \label{fig:transfer study}
\end{figure*}

See \Fref{fig:transfer study}.

\subsection{\Name's Extracting Mode.} 
\begin{figure*}[htbp]
  \centering
  \begin{subfigure}[b]{\textwidth}
  \centering
    \includegraphics[width=0.9\textwidth]{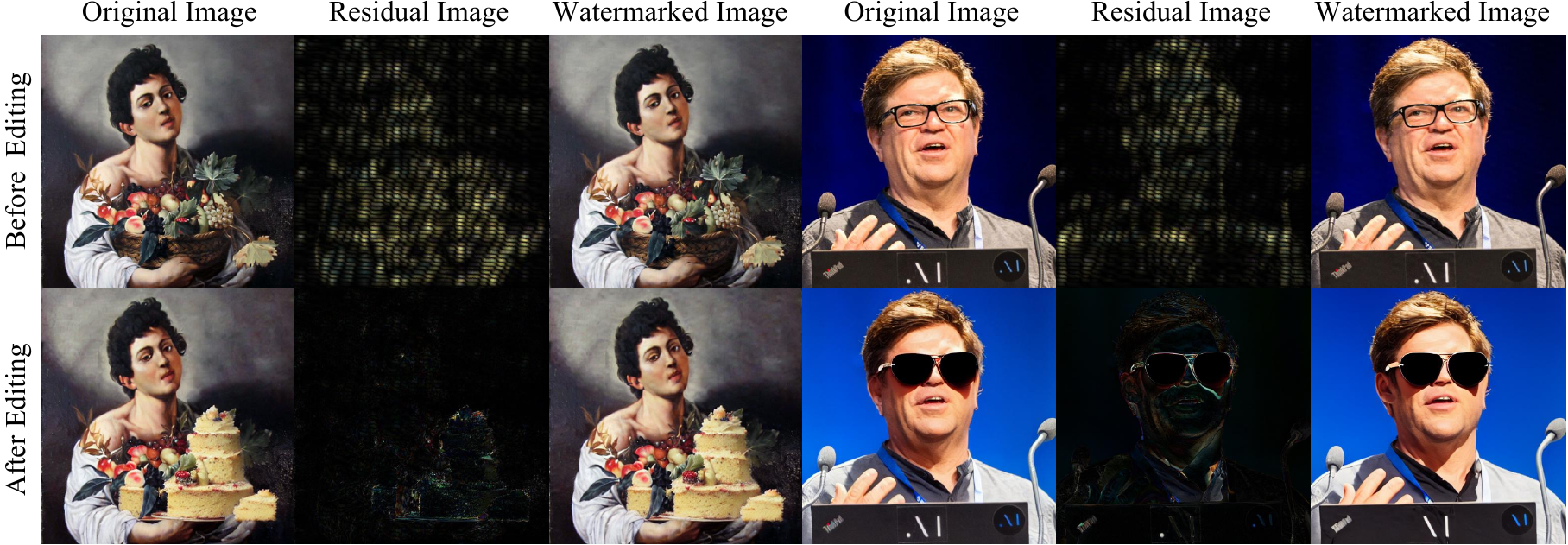}
    \label{fig:sub1}
  \end{subfigure}

  \vspace{1em}

  \begin{subfigure}[b]{\textwidth}
  \centering
    \includegraphics[width=0.9\textwidth]{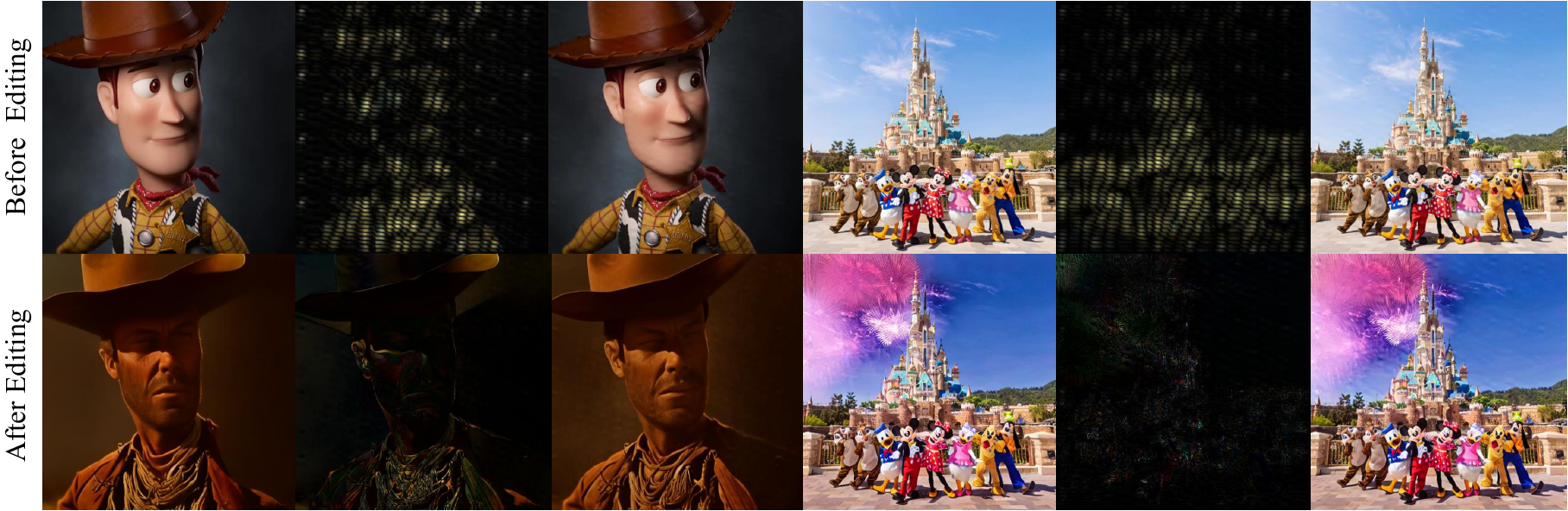}
    \label{fig:sub2}
  \end{subfigure}

  \vspace{1em}
    
  \begin{subfigure}[b]{\textwidth}
  \centering
    \includegraphics[width=0.9\textwidth]{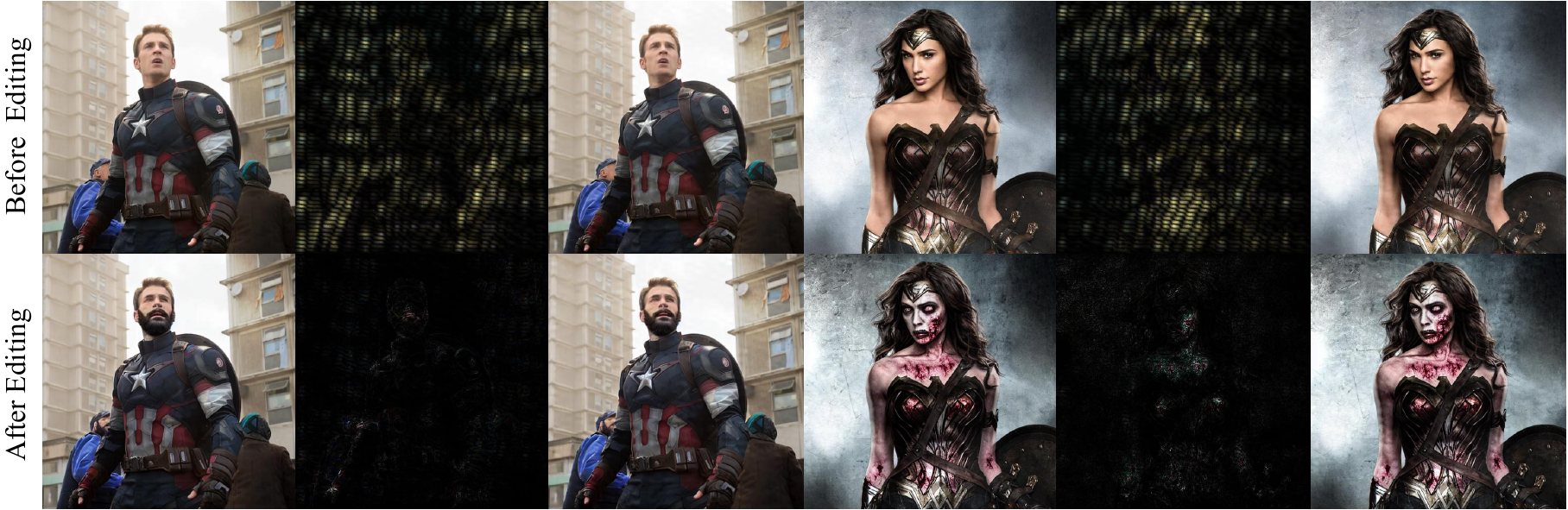}
    \label{fig:sub3}
  \end{subfigure}
  
  \caption{More visual examples of \Name's extracting mode.}
  \label{fig:extract patterns}
\end{figure*}

See \Fref{fig:extract patterns}.

